\documentclass[aps,prx,amsmath,twocolumn,groupedaddress,letterpaper,floatfix]{revtex4}
\usepackage{graphicx,color}
\usepackage{verbatim}
\usepackage{amssymb}   
\usepackage{amsmath}
\usepackage{amsfonts}
\usepackage{mathdots}
\usepackage{hyperref}
\usepackage{epsfig}
\usepackage{braket}
\usepackage{bm}
\begin{document}
	\title{ Topological superconductivity in EuS/Au/superconductor heterostructures}
	\author{Ying-Ming Xie$^1$}
	\author{K. T. Law$^1$} 
	\author{Patrick A. Lee$^2$}\thanks{Corresponding author.\\palee@mit.edu}
		\affiliation{$^1$Department of Physics, Hong Kong University of Science and Technology, Clear Water Bay, Hong Kong, China} 
	\affiliation{$^2$Department of Physics, Massachusetts Institute of Technology, Cambridge MA 02139, USA}

	\date{\today}
	\begin{abstract}
		In a recent work \cite{Manna2019}, signatures of a pair of Majorana bound states (MBS) were found in a new experimental platform formed by EuS islands deposited on top of a gold surface which was made superconducting through proximity coupling to a superconductor. In this work, we provide a theoretical understanding for how MBS can be formed in EuS/Au/superconductor heterostructures. We focus on the strip geometry where a narrow ferromagnetic strip is deposited on a planar structure. We first explicitly map out  the topological phase diagram of the EuS/Au/superconductor heterostructure using the lattice Green's function method. Importantly, we find that the chemical potential step between the region with and without EuS covering is a  crucial ingredient for the creation of MBS of this set-up. Next, we focus on the Bogoliugov quasi-particles that are bound to the region under the EuS by Andreev reflections from the surrounding superconductors. Moreover, we obtain the topological regimes analytically using the scattering matrix method. Notably, we confirm that the  normal reflections induced by the chemical potential step are essential for creating finite  topological regimes. Furthermore, the area of the topological regimes shows periodic oscillation as a function of chemical potential as well as the sample width. 
		We conclude by showing that the feromagnetic strip geometry holds a number of advantages over other quasi-one-dimensional schemes that have been proposed.
	\end{abstract}
	\maketitle
	\section{Introduction}
	Recently, there has been intense interest in creating Majorana bound states (MBS) in condensed matter systems. Of special interest are the MBS, which have been proposed to be building blocks of fault-tolerant quantum computers \cite{Kitaev1,Nayak2008}.  The MBS have been proposed to exist in the vortex cores of two-dimensional (2D) $p$-wave superconductors \cite{Read2000} or the ends of 1D $p$-wave  superconductors \cite{Kitaev2001}, where the topological superconductivity is formed. Recent efforts have focused on engineering structures where conventional superconductors can induce topological superconductivity via proximity effect \cite{Fu2008,Alicea2010}. Examples of these candidate topological superconductors 
	include superconductivity proximitized topological insulators \cite{Fu2008,Jia2015,Jia2016}, semiconductor nanowires \cite{Jaysau2010,Roman2010,Oreg2010,Patrick2010,Roman2011,Yingming2019,Mourik2012,Rokhinson2012,Das2012,Deng2012,Albrecht2016,Zhang2018,Marcus2020}, magnetic atom chains \cite{Choy2011,DanielLoss2013,NadjPerge2013,Pascal2017,Yazdani2014,Ruby2015,Pawlak2016,Kimeaar2018}, Majorana planar junctions \cite{Fornieri2019,Ren2019},   iron based superconductor FeTe$_{0.5}$Se$_{0.5}$\cite{Dinghong1,Dinghong2,Dinghong3},  a carbon nanotube \cite{Desjardins2019}  and higher order topological insulators \cite{Berthold}. However, finding an experimental platform which can easily scale up for creating and entangling a large number of MBS for quantum computation remains a major challenge. 
	
	\begin{figure}[ht]
		\centering
		\includegraphics[width=1\linewidth]{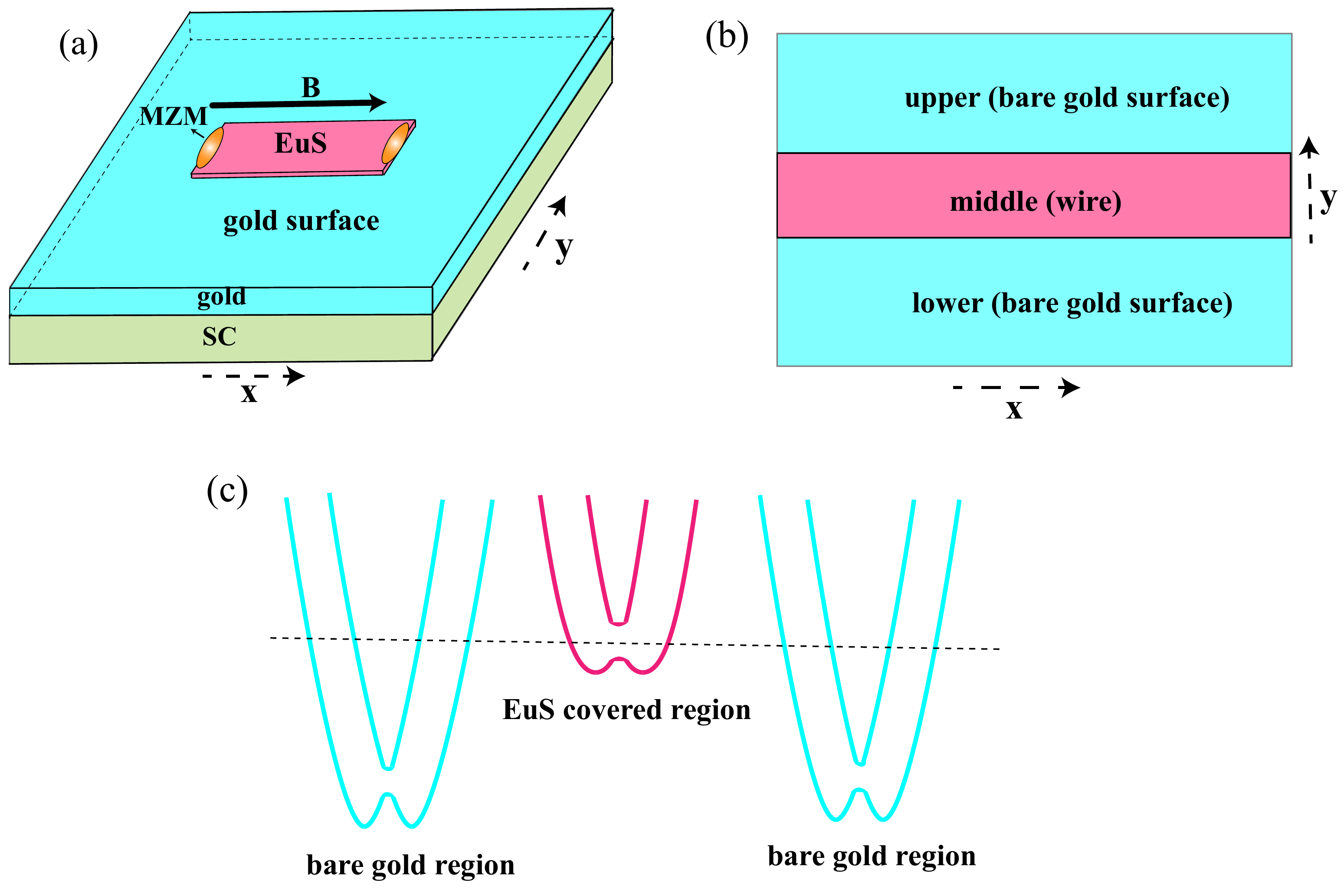}
		\caption{(a) The schematic figure of  EuS/Au/superconductor heterostructure used in ref.~\cite{Manna2019}. An EuS island is deposited on Au [111] surface which is in proximity to a parent superconductor. Upon applying an in-plane magnetic field $\bm{B}$, MBS appear at the ends of an EuS island. (b) The geometry employed in our calculation. The upper (U) and lower (L) regions are bare gold surfaces. The middle (M) region is the EuS covered gold surface forming a wire. We take the periodic boundary condition in $x$-direction  so that $k_x$ is a good quantum number and then take the infinite length limit. (c) The schematic picture of the band positions of gold surface states for the bare gold region and the EuS covered region. The dashed line indicates the position of the Fermi energy. }
		\label{fig:FIG1}
	\end{figure}
	
	Recently,  ferromagnetic EuS islands were deposited on  gold surfaces whose surface state has been made superconducting by the proximity effect.  With the application of  an in-plane magnetic field, zero-bias peaks were observed simultaneously at the two ends of the EuS islands \cite{Manna2019}. The observations were taken as evidence for the simultaneous appearance of MBS at opposite ends of a  topological superconductor \cite{Kitaev2001,Roman2010,Oreg2010,Patrick2010,Roman2011}, where pairs of MBS are separated spatially and topologically protected by the bulk superconducting gap.   In this paper we focus on the strip geometry where a ferromagnetic strip is deposited on the (111) surface of gold. A schematic experimental setup is shown in Fig.~\ref{fig:FIG1}(a). This setup is a further development of the original proposal of Potter and Lee \cite{Patrick2012}, which pointed out that the quasi-one-dimensional gold wires with [111] surface states can be used to realize topological superconductors. However, a fundamental difference  is that in the present setup, the gold is planar and only the ferromagnetic strip is quasi-one-dimensional. This difference requires a totally different understanding of the transverse confinement of the electrons and is one of the main focus of this paper. On the other hand, both schemes take advantage of the fact that the gold [111] surface state exhibits strong Rashba spin-orbit coupling (SOC) which causes a band splitting of about 110 meV and the SOC is several orders of magnitude larger than those in semiconductor nanowires \cite{Mourik2012}. The large Rashba SOC can ensure that the proximity superconducting pairing gap induced on the gold surface state is large even under a strong magnetic field.
	
	We note that magnetic islands or thin-flims coupled to conventional superconductors have also been  used to realize two-dimensional superconductivity  with chiral Majorana fermions \cite{Menard2017,Palacio2019,Menard2019,Garnier2019,Peter2020,Andrzej2020}, such as a monolayer of Pb covering magnetic Co–Si islands
	grown on Si(111) \cite{Menard2017} and nanoscale Fe islands on a Re surface \cite{Palacio2019}. However, the physics and the issues involved are quite different from the present setup, as will be discussed in more detail in the concluding section.
	
	The reason for switching to a ferromagnetic strip covered gold surface instead of using bare gold strips as originally proposed by Potter and Lee \cite{ Patrick2012} is that the original proposal has some limitations. First, the Fermi energy of the bare gold surface state is relatively high, roughly 500 meV above the band bottom of the surface Rashba band. As a result, many sub-bands will be partially occupied in a quasi-one-dimensional wire at the Fermi energy.  For example, roughly 100 sub-bands will be partially occupied if the gold wire is 100 nm wide \cite{Manna2019, Patrick2012}. This results in a large number of trivial end states co-existing with the MBS even in the topological regime \cite{Patrick2012}. Second, the g-factor of gold is about 2 which means that it requires a large external magnetic field to overcome a trivial pairing gap and re-open a topological gap. In the experiment, the proximity superconducting gap on gold using Vanadium is about 0.5 meV \cite{Patrick2019}. Therefore, it requires a magnetic field of about 10T to reach the topological regime which is experimentally difficult to achieve in an STM setting. Such a large magnetic field can also severely suppress the superconductivity in the parent superconductor. 
	
	Remarkably, it turns out that depositing EuS onto the gold surface solves the two aforementioned limitations at once. First, the surface Rashba band of the gold surface is shifted up so that the Fermi energy is only about 30 meV \cite{Manna2019,Patrick2019} above the band bottom. At the same time, EuS, being a ferromagnetic material, introduces a large exchange field which effectively enhances the Zeeman field \cite{Manna2019,miao2015spin} . With the EuS/Au/superconductor heterostructure geometry as shown in Fig.~\ref{fig:FIG1}(a), signatures of a pair of MBS appearing at the opposite ends of an elongated EuS island had been observed using STM measurements when a Zeeman field is applied along the island \cite{Manna2019}. 
	
	We emphasize an important difference  between the EuS strip setup and the bare gold strip setup\cite{ Patrick2012} in terms of the physics of the transverse confinement of the electrons. In the latter case the states are bound by the quantum well potential formed by the edges of of the gold strip, giving rise to a discrete set of transverse sub-bands. In the case of EuS strip the potential under the strip is higher, so that the electrons are repelled from the strip and the concept of transverse sub-bands does not apply. Instead, these electrons are Andreev reflected by the surrounding superconductors to form a bound state. As we shall see, both the Andreev reflections and normal backscatterings created by the chemical potential step are essential for giving rise to the topological regime.
	
	In ref.~\cite{Manna2019} numerical solutions were performed on  an effective tight-binding model to simulate the real-space features of MBS using realistic parameters. It leaves open the question as to how to optimize the parameters of this setup to obtain a robust topological superconductor. In this work, we probe deeper into the basic physics.   We perform more detailed numerical work and also provide analytical solutions to the model to bring more insight into the advantages of this setup, so that the question of how to optimize the topological gap can be answered. Our conclusion is that the ferromagnetic strip geometry holds a number of advantages over other schemes which have been proposed. A summary is is given in the conclusion section.
	
	This paper is organized as follows. In Section II, we calculate the topological phase diagram of an EuS strip deposited on a planar gold surface which is coupled to a superconductor. In the calculation,  we used relatively realistic parameters estimated from experiments \cite{Manna2019,LaShell} and map out the topological invariant of this inhomogeneity system using a lattice Green's function method. In particular, the self-energy renormalization effects and the spatial inhomogeneity of the electrostatic or chemical potential are  incorporated in our calculations (see Fig.~\ref{fig:FIG1}(c), the gold surface states in the EuS covered region and bare gold surface region possess different chemical potential).   We find that gold [111] surfaces with strips of EuS deposited exhibit sizable topological regimes and can be used to create Majorana fermions. Furthermore, we find that, in order to create MBS, it is essential to have a chemical potential step between the surface states covered by EuS and the bare gold surfaces. By gradually removing the chemical potential step, the topological regime diminishes and eventually vanishes.
	
	After the initial submission of this work, a paper by Papaj and Fu \cite{Papaj2021} appeared where they treated the  problem of a ferromagnetic insulator strip on top of a topological insulator. They obtained considerable insight to this problem by considering the Andreev and normal scattering of Bogoliubov quasiparticles by the boundary. Since the Rashba bands are essentially two copies of the surface states of a topological insulator with opposite helicity, we adopt the same method to obtain an analytic solution to our problem. This is discussed in section III. Similar to Papaj and Fu, to achieve a sizeable topological superconducting gap, the width of the strip should be comparable with the coherence length of the superconducting surface states.  On the other hand, unlike their problem, we find analytically that there is a periodic modulation of the topological regime induced by the chemical potential, in complete agreement with the numerical results.

	
	%

	
	\section{Topological superconductivity in EuS/Au/superconductor heterostructures: a numerical study} 
	
	\subsection{Model}Here, we study the topological properties of a ferromagnetic  magnetic material EuS island deposited on a gold surface which is coupled to a superconductor as depicted in Fig.~\ref{fig:FIG1}(a). We approach this problem by considering a sample shown in Fig.~\ref{fig:FIG1}(b) which has periodic boundary conditions in the  $x$-direction so that $k_x$ is a good quantum number, and infinite in the y-direction. The gold surface is separated into three segments, the upper bare gold surface region (U), the lower bare gold surface region (L) and the EuS covered gold surface in the middle region (M). We compute the topological invariant of this setup taking into account the 2D gold surface. 
	
	We first present the normal Hamiltonian that describes gold surface states. The continuum Hamiltonian that describes this partially covered gold surface state is 
	\begin{equation}
		H=\int dy \sum_{k_x}c^{\dagger}_{k_x,\alpha}(y)[h^{\alpha\beta}_{k_x}(y)+V(y)\sigma_{\alpha\beta}^x]c_{k_x,\beta}(y),
	\end{equation} 
	where
	\begin{equation}
		h_{k_x}(y)=\frac{k_x^2}{2m}-\frac{\partial_y^2}{2m}-\mu(y)+\alpha_R(k_x\sigma^y+i\partial_y\sigma^x).
	\end{equation}
	Here, $\sigma^{i}$ is the spin operator, $\alpha_R$ is the Rashba velocity characterizing the strength of spin-orbit coupling, $\mu(y)$ and $V(y)$ are the chemical potential  and the Zeeman energy respectively.
	
	The $y$ dependence of $\mu(y)$ and $V(y)$ captures the observation \cite{Manna2019,Patrick2019} that a thin layer of EuS can shift the chemical potential of the surface Rashba band so that the band bottom is moved from $500$ meV to around $30$ meV. At the same time, the Zeeman energy is locally enhanced under the ferromagnetic material EuS via the exchange coupling, which enables us to drive the gold surface states under the EuS island into the topological regime with a relatively small in-plane magnetic field.  
	
	We denote $\mu_1$, $V_1$ as the chemical potential and Zeeman energy for bare gold surface region where $y\in\{U,L\}$  and $\mu_2$ is the chemical potential for EuS covered gold surface where $y\in M$. 
	
	In the numerical calculations, we integrate out the bare gold regions numerically using lattice Green's function method \cite{Patrick1981,Sun_2009,Law2012,Law2013},  discretize the continuum Hamiltonian $H$ in the $y$-direction and obtain a lattice Hamiltonian $H_{0}$, where
	\begin{align}
		&H_{0}=\sum_{k_x,j}c^{\dagger}_{k_x,j,\alpha}((4t-\mu_j-2t\cos k_x)\delta_{\alpha\beta}+\alpha_R\sin k_x\sigma^y_{\alpha\beta}+\nonumber\\
		&V_j\sigma^{x}_{\alpha\beta})c_{k_x,j,\beta}+\sum_{k_x,j}c^{\dagger}_{k_x,j,\alpha}(-t\delta_{\alpha\beta}+\frac{i}{2}\alpha_R\sigma^{x}_{\alpha\beta})c_{k_x,j+1,\beta}+h.c..
	\end{align}  
	Here, we set $t=1/2ma^2=16$ eV$\cdot$\AA$^2/a^2$, $\alpha_R=0.4$ eV $\cdot$\AA$/a$, which are chosen to recover the realistic continuum band dispersion \cite{LaShell}. 
	%
	
	Next, we include  the superconductivity originating from the proximity effect into the model. In the EuS/Au/superconductor geometry, superconductivity is first induced on the gold bulk states through proximity effect. And the mixing of the gold bulk and surface states via impurity scattering or virtual scattering via phonon or Coulomb interaction can further induce superconductivity onto the surface states. As a result, the proximity effect on the surface states can be described  by a self-energy term \cite{Manna2019,Potter2011,Jay2010}
	\begin{equation}
		\Sigma(\omega^+)\approx-\Gamma \frac{(\omega^+-V_1\sigma^x)\tau_0-\Delta_B\tau_x}{\sqrt{\Delta_B^2-\omega_{+}^{2}}},
	\end{equation}
	where  $\omega_+=\omega+i\eta$, $\eta$ is an infinitesimal positive number,  the superconducting gap of gold bulk states $\Delta_B\approx0.5$ meV \cite{Manna2019}, $\tau$ operates on the Nambu particle-hole basis
	$\Psi(k_x,y)=[c_{k_x,\uparrow}(y),c_{k_x,\downarrow}(y),c^{\dagger}_{-k_x,\downarrow}(y),-c^{\dagger}_{-k_x,\uparrow}(y)]^{T}$, $\Gamma$ is the gold bulk and surface state mixing strength and is set to be 3$\Delta_B$ to explain the experimentally observed superconducting gap on the gold surfaces \cite{Manna2019}.   More specifically, $\Gamma = \pi N_{B}(0)W^{2}$ where $N_{B}(0)$ is the bulk density of states of gold near Fermi energy and $W$ is the disorder scattering strength which mixes the bulk and the surface states  \cite{Patrick2012}. Therefore, in our formalism, we take into account the effect of the coupling between the bulk states and the surface states of gold and do not use a simple Rashba band to describe the surface state. As we will see below, this indeed has an important effect on the localization length of the Majorana wavefunction of the system \cite{Pengyang}.
	
	After incorporating the self-energy term, the Green's function of the gold surface state  is	
	\begin{equation}
		G_0(\omega,k_x)=\frac{Z}{(\omega^+-V_{x}\sigma^{x})\tau_0-Zh_{k_x}\tau_z-(1-Z)\Delta_B\tau_x}.
		\label{Green_fun}
	\end{equation}
	The quasiparticle weight $Z(\omega_+)=\frac{1}{1+\Gamma/\sqrt{\Delta_B^2-\omega_{+}^{2}}}$. Here $V_x$ is an effective Zeeman energy. For the bare gold region, the effective Zeeman energy $V_x=V_1$ with $V_1=u_B B$ and $B$ is the strength of in-plane external field.  And for the EuS covered gold surface region, the effective Zeeman energy $V_x$ includes both $V_1$ and an additional Zeeman energy  induced by the exchange interaction of ferromagnetic material EuS, i.e., $V_x=ZV_{ex}+V_1$,  which for simplicity is replaced   by its zero frequency limit: $V_x\approx V_{EuS}$.
	
	\subsection{ Evaluating the $Z_2$ topological invariant}
	
	Our system breaks the time-reversal symmetry but preserves the particle-hole symmetry. As a result, the topological class of our model belongs to D class,  which is characterized by a $Z_2$ topological invariant \cite{Schnyder2008}. A simple scheme to obtain this $Z_2$ topological invariant for a quasi-one-dimensional system is to define a skew-symmetric matrix as $B(k_x)=H(k_x)\tau_y\sigma_y$ based on the particle-hole symmetry operator $\Theta=\tau_y\sigma_yK$,  and the $Z_2$ invariant $\mathcal{M}$ can be obtained as $\text{sgn}[\text{Pf} B(k_x=0)]\times \text{sgn}[\text{Pf} B(k_x=\pi/a)]$ \cite{Jaysau2010,Stanescu2013}, where $\text{Pf}$ denotes the Pfaffian of a matrix. Here $H(k_x)$ is the full Bogoliubov-de Gennes Hamiltonian to model the topological superconductor, $K$ denotes the complex conjugate operator. 
	
	However, we cannot directly apply this scheme to evaluate the topological invariant for two reasons. First, we have a frequency-dependence in the self-energy term; second, in order to treat a bare gold surface which is truly 2D, we cannot use $H(k_x)$ directly which describes a quasi-one-dimensional system. The first obstacle can be removed by using the Green's function scheme to evaluate the topological invariant. According to Ref.~\cite{Wangzong1, Wangzong2},   this scheme can be simplified to obtain the topological invariant from  the effective Hamiltonian, which is expressed in terms of Green's function at zero frequency: $-G^{-1}(\omega=0,k_x)$. The second obstacle can be overcome by integrating out the two bare gold segments to obtain the self-energy terms  $\Sigma_{U}(\omega,k_x)$ and  $\Sigma_{L}(\omega,k_x)$ which can be added to the Green's function of gold surface covered by EuS. With Dyson's equation,
	$G(\omega,k_x)=(G^{-1}_0(\omega,k_x)-\Sigma_U(\omega,k_x)-\Sigma_L(\omega,k_x))^{-1}$, the effective Hamiltonian is obtained as
	\begin{align}
		h_t(k_x)=&h_{k_x}(y\in M)\tau_z+Z(0)^{-1}V_{EuS}\sigma_x+\nonumber\\
		&(Z(0)^{-1}-1)\Delta_B\tau_x+\Sigma_U(0,k_x)+\Sigma_L(0,k_x).\label{eff_Hamil}
	\end{align} 
	$\Sigma_{U(L)}(0,k_x)$ can be calculated from Eq.~\ref{Green_fun} numerically using the lattice Green's  function method \cite{Patrick1981,Sun_2009,Law2012,Law2013}. More details can be found in Supplementary Material \cite{Supp}. The  $B(k_x)$ can be defined as $h_t(k_x)\tau_y\sigma_y$, and this new skew-symmetric matrix is used to evaluate the topological invariant $\mathcal{M}$ for our model. Note that we take into account the fact that the relatively small Zeeman energy ($\sim0.2\Delta_B$) in bare gold surfaces cannot close the superconducting gap. This enables the bare gold region to be integrated out without introducing extra singularities into the Green's function. 
	
	\begin{figure}
		\centering
		\includegraphics[width=1\linewidth]{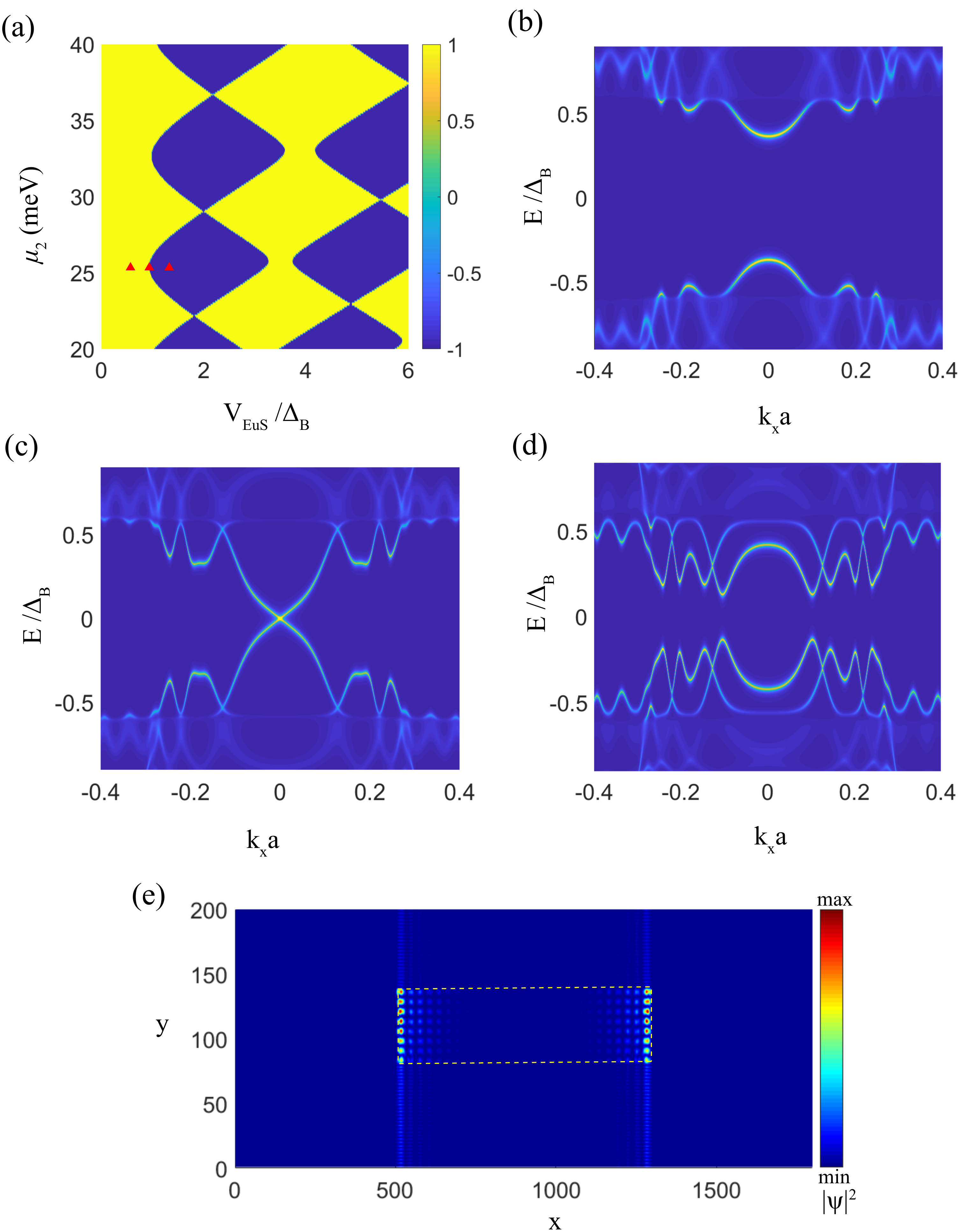}
		\caption{The topological phase diagram of the heterostructure formed by a 60 nm wide EuS strip and a large gold surface (2D limit here). The width of EuS strip is set to be 60nm. The parameter $\mu_1=500$ meV, $V_1=0.2\Delta_B$ for bare gold surface  are adopted. (a)  the topological invariant $\mathcal{M}$ is shown  as a function of effective Zeeman energy $V_{EuS}$ and chemical potential $\mu_2$. The topological trivial region with $\mathcal{M}=1$ is shown in yellow and topological nontrivial region with $\mathcal{M}=-1$ is shown in blue.  (b),(c),(d) show the spectral function $A(k_x,E)=-\text{Im}(\text{Tr}(G(\omega,\bm{k_x})))$ as a function a $k_x$ and $E$ with the parameters at red triangles shown in (a). The (b),(c),(d) shows the typical excitation feature at the  trivial region, phase transition boundary, and topological regime respectively.   The chemical potential $\mu_2$ of (b),(c), (d) are all chosen to be 25 meV. The $V_{EuS}$ equals  $0.5\Delta_B$ for (b), $0.95\Delta_B$ for (c), and $1.5\Delta_B$ for (d). (e) The Majorana wavefunction of a 800 nm $\times$ 60 nm  EuS (denoted by the yellow box) deposited on a 2000 nm $\times$ 200 nm Au surface with the parameters of (d). The color indicates the absolute value square of the MBS wavefunction. Here $a$ is set to be 1 nm to reduce the finite size effect.} 
		\label{fig:FIG2}
	\end{figure}
	\subsection{Phase diagram}
	In ref.~\cite{Manna2019},  signatures of a pair of MBS were observed when a EuS island was placed on a gold wire which was in proximity to a superconductor.  Here we show how the EuS/Au/superconductor heterostructure can become  a topological superconductor. To model the topological regime of a large gold surface case, we consider a heterostructure formed by covering a 60 nm wide EuS strip in the middle of a 2D gold surface. Following the scheme of evaluating $Z_2$ topological invariant shown in the previous section, the resulting phase diagram is obtained as Fig.~\ref{fig:FIG2}(a).  It is interesting to note that the phase diagram in Fig.~\ref{fig:FIG2}(a) resembles the phase diagram of superconducting quasi-one-dimensional gold wires subject to a Zeeman field. However, the physical origins of the topological regimes are very different. For quasi-one-dimensional gold wires, the system is topological when superconductivity is induced on a wire with an odd number of sub-bands partially occupied at the Fermi energy. In our current situation, the gold surface is strictly 2D and quasi-one-dimensional sub-bands are not well defined. On the other hand, the Zeeman field induced by the external magnetic field and the ferromagnetic material EuS can create in-gap  Andreev bound states under the EuS island \cite{Choy2011, Pascal2017}. As we will discuss later, these in-gap bound states are confined to be under the EuS island by the fully gapped gold surface states.  
	
	This phase diagram is further demonstrated with Fig.~\ref{fig:FIG2}(b),(c),(d). As the parameter, such as the effective Zeeman energy in this case, is tuned across the phase boundary, the energy gap closes and reopens which signals the topological phase transition. The in-gap quasi-particle bound states are clear in Fig.~\ref{fig:FIG2}(b),(c),(d). We emphasize these in-gap quasi-particle bound states are trapped under EuS covered region through both the Andreev reflection introduced by the gapped superconducting gold surface and the normal reflection of chemical potential step. More importantly, as shown in Fig.~\ref{fig:FIG2}(d), there is a relatively uniform and sizable topological gap ($\sim0.1 \Delta_B$) deep in the topological regime (for example, far away from the topological phase transition boundaries). The MBS using the parameters in Fig.~\ref{fig:FIG2}(d) is shown in Fig.~\ref{fig:FIG2}(e). MBS residing at the two ends of the EuS island can be clearly observed.  Notably, due to the in presence of self-energy term $\Sigma$ introduced in Eq.5, the localization length of the Majorana mode is  shorter than the estimated  superconducting coherence length of gold surface states $\xi_0$ which is $\approx t/\Delta_B \approx$ 320nm (cf. \cite{Manna2019, Pengyang} and Supplementary Material \cite{Supp} for more details). This is consistent with the short localization length (only tens of nm~) of the Majorana modes observed in the experiment \cite{Manna2019}.

	%
	
	\subsection{The importance of the chemical potential step}\label{sec_D}
\begin{figure}
		\centering
		\includegraphics[width=1\linewidth]{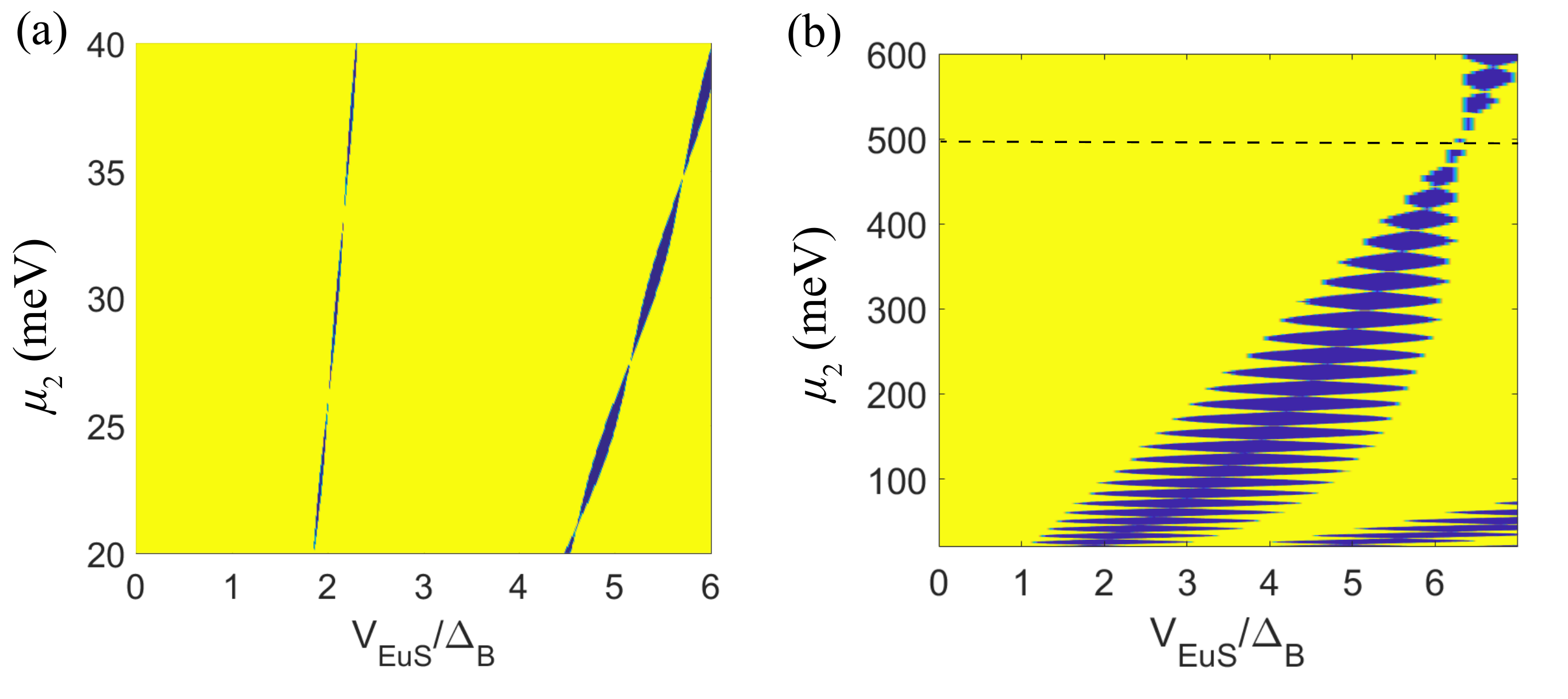}
		\caption{The effect of chemical potential step between the bare gold surface and the EuS covered region. (a) shows the phase diagram of the heterostructure formed by the EuS strip and the gold surface. The parameters are the same as Fig.~\ref{fig:FIG2}(a), except the chemical potential of the bare gold region artificially set as $\mu_1=\mu_2$. The yellow and blue areas represent the topological trivial and nontrivial regimes respectively. (b) The phase diagram as a function of Zeeman energy and the chemical potential of the ferromagnetic material covered region.   Here, the chemical potential of the bare gold is fixed at $\mu_1 = 500$ meV for (b), and $a$ is set to be $4$\AA\   to properly capture the dispersion of gold surface states near 500 meV. }
		\label{fig:FIG3}
	\end{figure}

	In the EuS/Au/superconductor heterostructure with EuS islands deposited on a 2D gold surface, there is a chemical potential step between the area under EuS and the bare gold surface. As shown experimentally, the chemical potential shift indeed depends on the thickness of EuS. When bilayer EuS is deposited on the gold surface, the chemical potential is shifted from $\mu_1\sim500$ meV to $\mu_2\sim 30$ meV relative to the surface Rashba band bottom \cite{Manna2019}. On the other hand, if a monolayer EuS is used, the chemical potential is shifted to about $200$meV instead \cite{Patrick2019}. In this section, we study the importance of this chemical potential step. First of all, if we remove this chemical potential step artificially by setting $\mu_1=\mu_2$, the phase diagram will change from Fig.~\ref{fig:FIG2}(a) into Fig.~\ref{fig:FIG3}(a). Surprisingly, the topological regimes  (in blue) become hardly visible, even though the chemical potential is very low. This implies not only the inhomogeneity of Zeeman energy but also the inhomogeneity  of chemical potential  is important for the observation of a sizable topological regime on a gold surface. It can be seen from Fig.~\ref{fig:FIG2}(a) that the separation between the diamond topological regimes is roughly 6 meV, which is  expected for a wire of the width of the EuS (see Supplementary Material \cite{Supp}). This suggests the chemical potential step effectively creates a sample width given by EuS width due to the scattering from the potential step. In the calculation, we used a step function to  describe the chemical potential shift induced by EuS, although  from microscopic point of view, the chemical potential transition region may extend over several lattice constant.  This approximation should be valid as long as the length of transition region is much smaller than the size of islands and gold surface.
	
	On the other hand, if EuS with a different thickness or other ferromagnetic materials  are deposited on the gold surface, the shift in chemical potential can be different. In Fig.~\ref{fig:FIG3}(b), we calculated the topological regime with a wide range of chemical potential underneath the ferromagnetic material, using the parameters of Fig.~\ref{fig:FIG2}(a) except the range of chemical potential used. It is clear from Fig.~\ref{fig:FIG3}(b) that a sizable chemical step between the area covered by the ferromagnetic material and the bare gold surface is needed to create large topological regimes. 
	
	\section{Topological regimes of a magnetic strip/Rashba superconductor heterostructure: an analytical study using the scattering matrix method}
	In the previous part, we have explicitly mapped out  the topological phase diagram of the EuS/Au/superconductor heterostructure using the lattice Green's function method. The features of in-gap bound states and the importance of chemical potential steps are recognized in this topological heterostructure.  In this section, we treat this system as a magnetic strip/Rashba superconductor junction with a uniform superconducting phase and determine  the topological regimes  analytically by solving the energies of Andreev bound states using the scattering matrix method \cite{Beenakker1991,Pientka2017, Papaj2021}. 
	
	\begin{figure}
		\centering
		\includegraphics[width=1\linewidth]{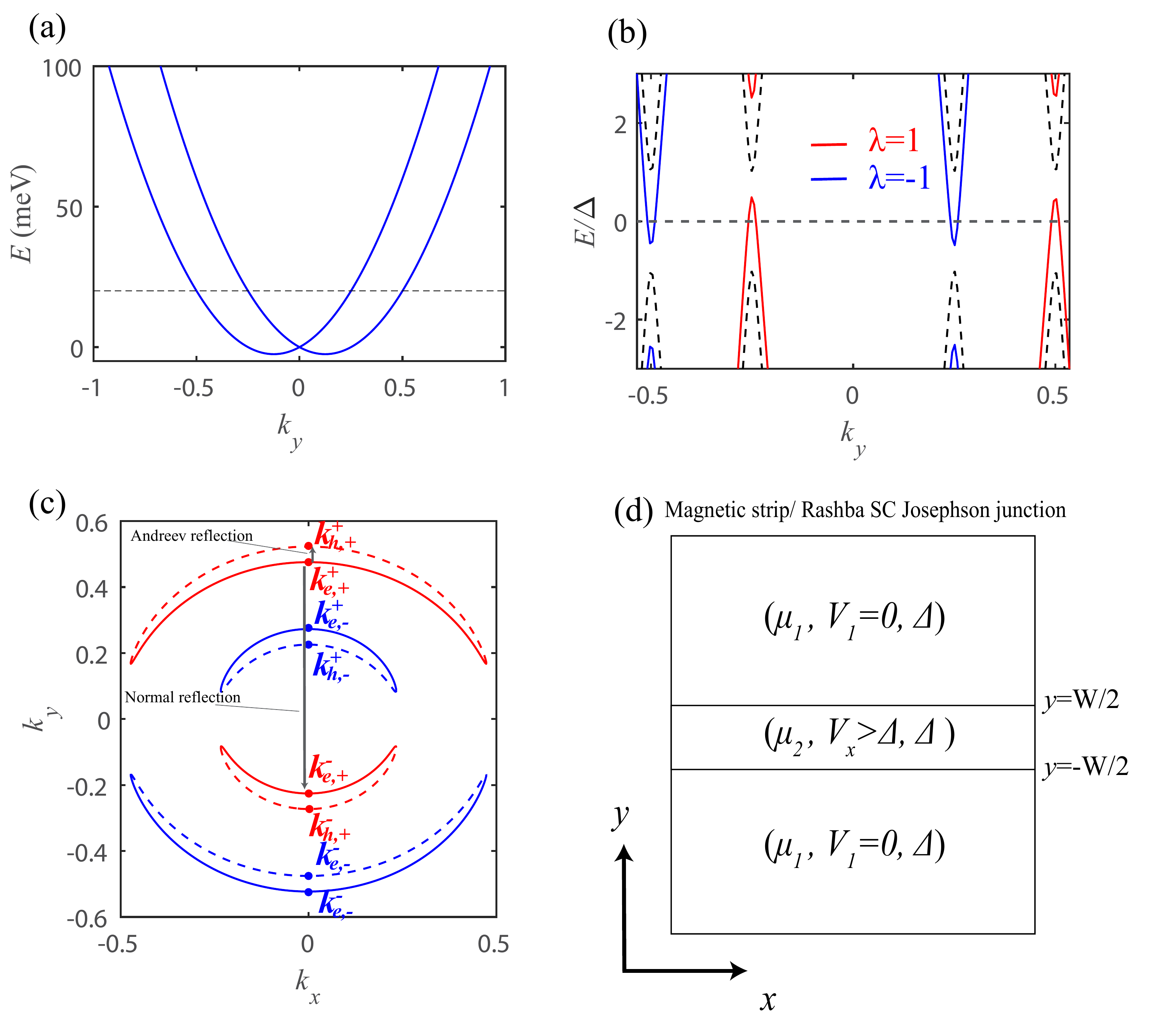}
		\caption{(a)The Rashba band in the normal state along $k_y$. Dotted line marks the chemical potential. (b) The BdG spectrum from Eq.~(\ref{model_Ha}) along $\bm{k}=(k_x=0,k_y)$. The black dashed lines are the BdG spectrum for $V_x=0$, while the colored lines  are the BdG spectrum for $V_x/\Delta=1.5$ $\lambda=1$ in red, $\lambda=-1$ in blue denotes state with spin parallel or anti-parallel to the Zeeman field. Note that the spectrum become gapless when $V_x>\Delta$. (c) shows the energy contours of the BdG spectrum at $E=0$ with $V_x/\Delta=2$, where electron-dominant  and hole-dominant arcs (the corresponding k states are labeled with the subscript $e$ and $h$) are highlighted as solid and dashed lines, respectively. With $\nu$ labeling the sign of $k_y$, the 8 momenta with the label $k_{e(h),\lambda}^{\nu}$  given by Eq.~(\ref{wave_vec}) are shown and easily visualized. A normal reflection between inter-Rashba Fermi circle and an Andreev reflection  within intra-Rashba Fermi circle  are highlighted. For $k_x=0$ the red and blue momentum states do not admix. (d) A  schematic plot of the geometry of the magnetic strip/Rashba superconductor junction considered in our derivation, where a magnetic trip partially covers a planar Rashba superconductor. The chemical potential, Zeeman energy ($\mu$, $V_x$, $\Delta$) of  different regions are highlighted.    }
		\label{fig:fig4}
	\end{figure}
	\subsection{Origin of the Andreev bound states   in a magnetic strip/Rashba superconductor junction}
	Let us first illustrate the origin of the Andreev bound states   in the magnetic strip/Rashba superconductor junction. We start from the following BdG Hamiltonian:
	\begin{equation}
		H(\bm{k})= [\xi_{\bm{k}}+\alpha_R (k_x\sigma_y-k_y\sigma_x)]\tau_z+V_x\sigma_x+\Delta\tau_x, \label{model_Ha}
	\end{equation}
	where  the Hamiltonian is  defined in Nambu basis $(c_{\bm{k},\uparrow}, c_{\bm{k},\downarrow}, c^{\dagger}_{-\bm{k},\downarrow},- c^{\dagger}_{-\bm{k},\uparrow})^{T}$,   $\sigma_i$ and $\tau_i$, respectively, operate on the spin and particle-hole space, the kinetic energy term $\xi_{\bm{k}}=\bm{k}^2/2m-\mu$, $V_x$ is the Zeeman energy, and $\Delta$ is the pairing potential. In this section, the proximity effects from the magnetic strip and the parent superconductor are responsible for inducing the Zeeman term and  pairing term in Eq.~(\ref{model_Ha}). For simplicity, we neglect the Zeeman energy $V_1$ in the bare superconducting region.  To clearly show the origin of the Andreev bound states, we display the energy dispersion of normal states in Fig.~\ref{fig:fig4}(a) and the BdG spectrum from $H(\bm{k})$ at $V_x/\Delta=0$ (blacked dashed line) and $V_x/\Delta=1.5$ (red and blue) in Fig.~\ref{fig:fig4}(b). In the absence of the Zeeman energy $V_x$, a superconducting gap of $\Delta$ is opened near Fermi energy, while a finite Zeeman term would suppress the excitation gap. When the Zeeman energy exceeds the pairing potential, i.e., $V_x>\Delta$, the excitation spectrum becomes gapless (see Fig.~\ref{fig:fig4}(b)). These gapless excitations  caused by the large Zeeman energy from magnetic strip  would result in some Fermi contours at $E=0$,  as shown in Fig.~\ref{fig:fig4}(c). Each contour consists of an electron-dominated and a hole-dominated arc. In contrast, as depicted in Fig.~\ref{fig:fig4}(d)  the bare superconducting regions still  possess a large superconducting gap $\Delta$,   thereby confining those in-gap excitations under the magnetic strip  as Andreev bound states. 
	\subsection{Topological phase transition boundaries using scattering matrix method}
	The boundaries of topological phase transitions are determined by $\epsilon(k_x=0)=0$ with $\epsilon(k_x)$ as the energy of Andreev bound states.    Here, we chose the strip to be along $x$-direction so that $k_x$ is a quantum number to label the states.  When $k_x=0$, the model Hamiltonian Eq.~(\ref{model_Ha}) becomes 
	$
	H(\bm{k})=(\xi_{\bm{k}}-\alpha_R k_y\sigma_x)\tau_z+V_x\sigma_x+\Delta\tau_x$.
	In this case, the model Hamiltonian $H(\bm{k})$ exhibits a chiral symmetry $[\sigma_x, H(\bm{k})]=0$. 	Thus, we can choose the spin quantization axis along $x$-direction, and block diagonalized the Hamiltonian as: 
	\begin{equation}
		H(\bm{k})=\begin{pmatrix}
			H_{+}(\bm{k})&0\\0&H_{-}(\bm{k})
		\end{pmatrix}.
	\end{equation}
	where $H_{\lambda}=(\xi_{\bm{k}}-\lambda\alpha_R k_y)\tau_z+\lambda V_x+\Delta \tau_x$. The BdG spectrum of  $\lambda=1$ and $\lambda=-1$ block are highlighted as red and blue color in Fig.~\ref{fig:fig4}(b). These are simply spin polarized states which are parallel or anti-parallel to the Zeeman field. As these two blocks do not mix, we can solve the bound states given by two blocks separately.
	
	Next, let us solve the  Andreev bound states in this magnetic strip/Rashba superconductor junction using the scattering matrix method. The first step is to solve the eigenmodes of different parts of the junction, where the chemcial potential, Zeeman term, and  pairing potential are labeled explicitly in Fig.~\ref{fig:fig4}(d). Note that if $\mu_1\neq \mu_2$, it indicates a chemical potential step. In the middle region where the magnetic strip covers with $V_x>\Delta$, the eigenstate is
	\begin{equation}
		\psi_{\beta,\lambda}^{\nu}(y)=\sqrt{\frac{\Delta}{2V_x}}\begin{pmatrix}
			e^{-\frac{1}{2}\rho^{\beta}\text{acosh}\frac{V_x}{\Delta}}\\
			\lambda	e^{\frac{1}{2}\rho^{\beta}\text{acosh}\frac{V_x}{\Delta}}
		\end{pmatrix}e^{ik_{\beta,\lambda}^{\nu}y},
	\end{equation}
	where  the bound state energy $\epsilon_{\lambda}(k_x=0)=0$ is considered, $\beta=e,h$ labels the electron/hole-dominated mode and $\rho^{e/h}=1/-1$. We introduce  the wavevectors 
	\begin{eqnarray}
		k_{\beta,\lambda}^{\nu}&&=k_{F,\lambda}^{\nu}+\lambda\rho^{\beta}\nu\frac{m\sqrt{V_x^2-\Delta^2}}{\sqrt{m^2\alpha_R^2+2m\mu_2}},\label{wave_vec}\\ k_{F,\lambda}^{\nu}&&=\lambda m\alpha_R+\nu\sqrt{m^2\alpha_R^2+2m\mu_2}.\label{momentum_middle}
	\end{eqnarray}
	Here $\nu=1/-1$ labels the positive/negative wavevector. The eight possible $k^{\nu}_{\beta,\lambda}$ wavevectors are highlighted in Fig.~\ref{fig:fig4}(c). Similarly, 
	in the top and bottom bare superconducting region ($V_x=0$), the eigenstates with $\epsilon_{\lambda}(k_x=0)=0$  becomes
	\begin{equation}
		\psi'_{\beta,\nu}(y)=\frac{1}{\sqrt{2}}\begin{pmatrix}
			1\\
			i\rho^{\beta}
		\end{pmatrix}e^{ik'^{\nu}_{\beta,\lambda}y},
	\end{equation}
	where the wavevectors
	\begin{eqnarray}
		k'^{\nu}_{\beta,\lambda}=k'^{\nu}_{F,\lambda}+\nu\rho^{\beta}\frac{i\Delta}{\sqrt{m^2\alpha_R^2+2m\mu_1}},\\ k'^{\nu}_{F,\lambda}=\lambda m\alpha_R+\nu\sqrt{m^2\alpha_R^2+2m\mu_1}.
	\end{eqnarray}
	
	Next, we employ the continuity of the wavefunction and the conversation of the probability current to obtain the condition for the appearance of  $\epsilon(k_x=0)=0$, where a gap closing appears at $k_x=0$ and would indicate a topological phase transition. Instead of matching the boundary conditions for the four waves, it is advantageous to  use the scattering matrix method \cite{Beenakker1991} which investigates the relation between the incoming states $\psi^{in}=(a_{e,\lambda}^{-}(L),b^+_{h,\lambda}(L), a^{+}_{e,\lambda}(U), b_{h,\lambda}^{-}(U))^{T}$ and the outgoing state $ \psi^{out}=(a_{e,\lambda}^{+}(L),b^-_{h,\lambda}(L), a^{-}_{e,\lambda}(U), b_{h,\lambda}^{+}(U))^{T}$ with the wavefunction in the middle region  decomposed as $\psi(y)=\sum_{\nu,\lambda}a_{e,\lambda}^{\nu}\psi_{e,\lambda}^\nu(y)+b_{h,\lambda}^{\nu}\psi_{h,\lambda}^\nu(y)$ (see Supplementary Material \cite{Supp} for the details). 
	
	On one hand, the incoming state will be scattered as the outgoing states at the interfaces $|y|=W/2$, i.e., $\psi^{out}=S\psi^{in}$. Here,  $W$ is the width of the junction,  the scattering matrix $S=[S_L,0;0,S_U]$ with $S_{U(L)}$ as the scattering matrix at upper (lower) interface at $y=W/2$ ($y=-W/2$).  On the other hand, the outgoing states will be transmitted as incoming states during the propagation within the middle region, i.e., $\psi^{in}=T\psi^{out}$ with the transition matrix $T=[0, T_{LU};T_{UL},0]$. The combination of $\psi^{out}=S\psi^{in}$ and $\psi^{in}=T\psi^{out}$ requires 
	$\det{[I-ST]}=1$ with $I=\text{diag}(\mathbb{I},\mathbb{I})$,  which gives
	\begin{equation}
		\det[\mathbb{I}-S_UT_{UL}S_LT_{LU}]=0. \label{Eq:bc}
	\end{equation} 
	After some explicit derivations (see Supplementary Material \cite{Supp} for the details), we found that the scattering matrices can be expressed as
	
	\begin{equation}
		S_L=S_U=\begin{pmatrix}
			i\lambda r e^{i\phi_{\lambda}}&-\sqrt{1-r^2}e^{i\phi_{\lambda}}\\
			-\sqrt{1-r^2}e^{i\phi_{\lambda}}&	i\lambda r e^{i\phi_{\lambda}}
		\end{pmatrix}\equiv\begin{pmatrix}
			r_{e}&r_{A}\\
			r_A&r_{h}	
		\end{pmatrix}.\label{scatter_ma}
	\end{equation}
	with
	\begin{widetext}
		\begin{eqnarray}
			r_e=r_h=\frac{(\mu_1-\mu_2)\sinh\gamma}{-i\lambda(m\alpha_R^2+\mu_1+\mu_2)\sinh\gamma+\sqrt{(m\alpha_R^2+2\mu_1)(m\alpha_R^2+2\mu_2)}};\label{normal_reflection}\\
			r_A=\frac{\ \sqrt{(m\alpha_R^2+2\mu_1)(m\alpha_R^2+2\mu_2)}\text{cosh}\gamma}{-i\lambda(m\alpha_R^2+\mu_1+\mu_2)\sinh\gamma+\sqrt{(m\alpha_R^2+2\mu_1)(m\alpha_R^2+2\mu_2)}}.
		\end{eqnarray}
	\end{widetext}
	where $\gamma=\text{acosh}\frac{V_x}{\Delta}$, $r_A$ is  from Andreev reflections, and $r_{e(h)}$ is from  normal reflections being  finite when $\mu_1\neq \mu_2$. As highlight in Fig.~\ref{fig:fig4}(c), due to the spin-orbit locking,  $r_A$ is induced by the scattering between an electron-dominant and a hole-dominant  arc  from intra-Rashba Fermi circle, while $r_e(h)$ is induced by the scattering between two electron (hole)-dominant arc  from inter- Rashba Fermi circle.
	The transmission matrices $T_{LU}$ and $T_{UL}$ are expressed as 
	\begin{equation}
		T_{LU}=\begin{pmatrix}
			e^{-ik^{-}_{e,\lambda}W}&0\\
			0&e^{-ik^{+}_{h,\lambda}W}
		\end{pmatrix}, 	T_{UL}=\begin{pmatrix}
			e^{ik^{+}_{e,\lambda}W}&0\\
			0&e^{ik^{-}_{h,\lambda}W}
		\end{pmatrix}.\label{trans_ma}
	\end{equation}
	Inserting Eq.~(\ref{scatter_ma}) and Eq.~(\ref{trans_ma}) back to  Eq.~(\ref{Eq:bc}), we find the gap closes at $k_x=0$  when 
	\begin{equation}
		r^2\cos(2\sqrt{m^2\alpha_R^2+2m\mu_2}W)+\cos(2\lambda\theta W-2\phi_{\lambda})=1-r^2,\label{gapclose}
	\end{equation}
	where $\theta=m\sqrt{V_x^2-\Delta^2}/\sqrt{m^2\alpha_R^2+2m\mu_2}$. Eq.~(\ref{gapclose}) is the central result of this section.
	
\begin{figure*}
		\centering
		\includegraphics[width=1\linewidth]{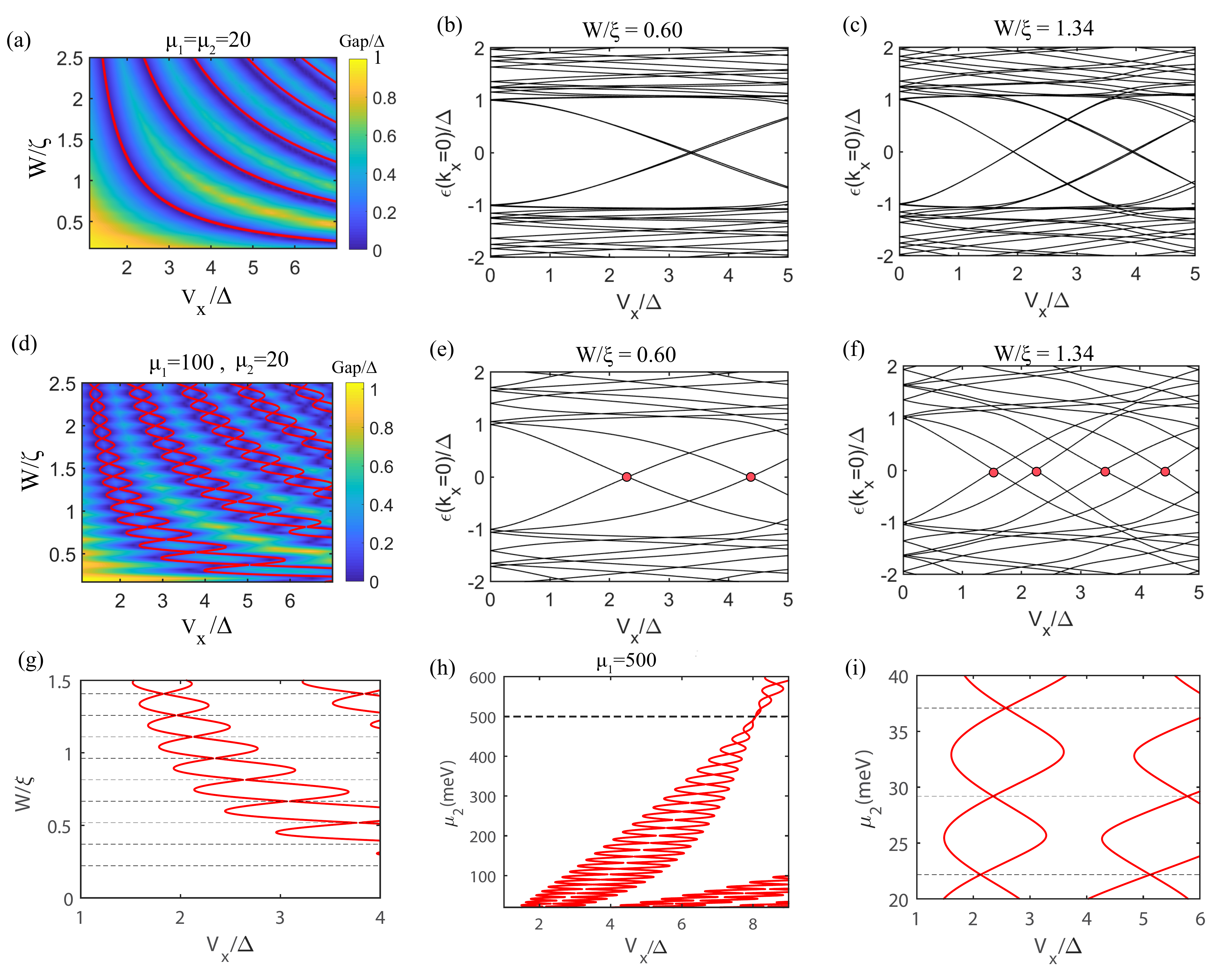}
		\caption{(a) and (d) show the topography of excitation gap at $k_x=0$ as a function of junction width $W/\xi$ and Zeeman energy $V_x/\Delta$  without chemical potential steps ($\mu_1=\mu_2=20$ meV) and with  a chemical potential step ($\mu_1=100,\mu_2=20$ meV), respectively. The red lines are the gap closing lines  indicated by our analytical solution Eq.~(\ref{gapclose}). (b), (c) and (e), (f) show  excitation spectrum $\epsilon(k_x=0)$ vs $V_x$ at $W/\xi=0.60$, $W/\xi=1.34$ for the case without chemical potential steps ($\mu_1=\mu_2=20$ meV) and with  a chemical potential step ($\mu_1=100, \mu_2=20$ meV),  respectively. The red dots in (e) and (f) mark the gap closing points which set the boundaries of the topological regime in (d). The similar pair of gap closing points sits on top of each other in (b) and (c) and is not shown.  (g) An enlargement of (d), showing only the analytic result.  The black dashed lines indicate the periodicity of the oscillations of topological regimes given by $\cos(2\sqrt{m^2\alpha_R^2+2m\mu_2}W)=-1$. (h) shows the gap closing from Eq.~(\ref{gapclose}) as a function of $\mu_2$ and $V_x$, where $\mu_1=500$ meV, and the width $W/\xi=1.06$. (i) is a zoomed-in version of (h). The additional black  dashed lines are from $\cos(2\sqrt{m^2\alpha_R^2+2m\mu_2}W)=-1$. }
		\label{fig:fig5}
	\end{figure*}

	We first consider  the case without  chemical potential step, i.e., $\mu_1=\mu_2$, so that the normal reflection vanishes $r=0$ (see Eq.~(\ref{normal_reflection})). In this case, the gap closes along a single line given by
	\begin{equation}
		\frac{W}{\xi}=\frac{\lambda(\phi_{\lambda}+n\pi)\sqrt{1+\frac{m\alpha_R^2}{2\mu_2}}}{\sqrt{(\frac{V_x}{\Delta})^2-1}}.\label{Eq:ana2}
	\end{equation}
	Here $n$ is an integer number, $\phi_{\lambda}=\text{Arg}[\frac{\cosh\gamma}{1-i\lambda\sinh \gamma}]$ and the width is naturally written as the dimensionless ratio $W/\xi$ where coherence length is defined as $\xi=v_{f2}/\Delta$ with $v_{f2}=\sqrt{2u_2/m}$. Noted that here we used $\mu_2$ in the Fermi velocity  instead of $\mu_1$, since the superconducting topological gap is dominant  by the coherence length characterized by $\mu_2$ within the junction instead of $\mu_1$ in the bare superconductor region.
	The topological regime actually vanishes in this case without the chemical potential step, in agreement with the numerical results, as shown in Fig.~\ref{fig:fig5}(a). 
	
	\subsection{Finite and periodically oscillating  topological regimes induced by the chemical potential step}
	To verify our analytical result Eq.~(\ref{gapclose}),  we calculate the energy gap at $k_x=0$ numerically by diagonalizing the tight-binding model as given in the Supplementary Material \cite{Supp}.  The gap at $k_x=0$ as a function of $W/\xi$ and $V_x/\Delta$ from Eq.~(\ref{gapclose}) are depicted in Fig.~\ref{fig:fig5}(a) without chemical potential steps ($\mu_1=\mu_2=20$ meV) and Fig.~\ref{fig:fig5}(d) with a chemical potential step ($\mu_1=100$ meV, $\mu_2=20$ meV).  The red solid lines correspond gap closing lines given by  Eq.~(\ref{gapclose}).  It can be seen that our analytical result matches  with the numerical result very well.  A small deviation is seen at large $V_x$ due to the violation of the assumption of $\mu\gg V_x$. 
	
	\begin{figure*}
		\centering
		\includegraphics[width=1\linewidth]{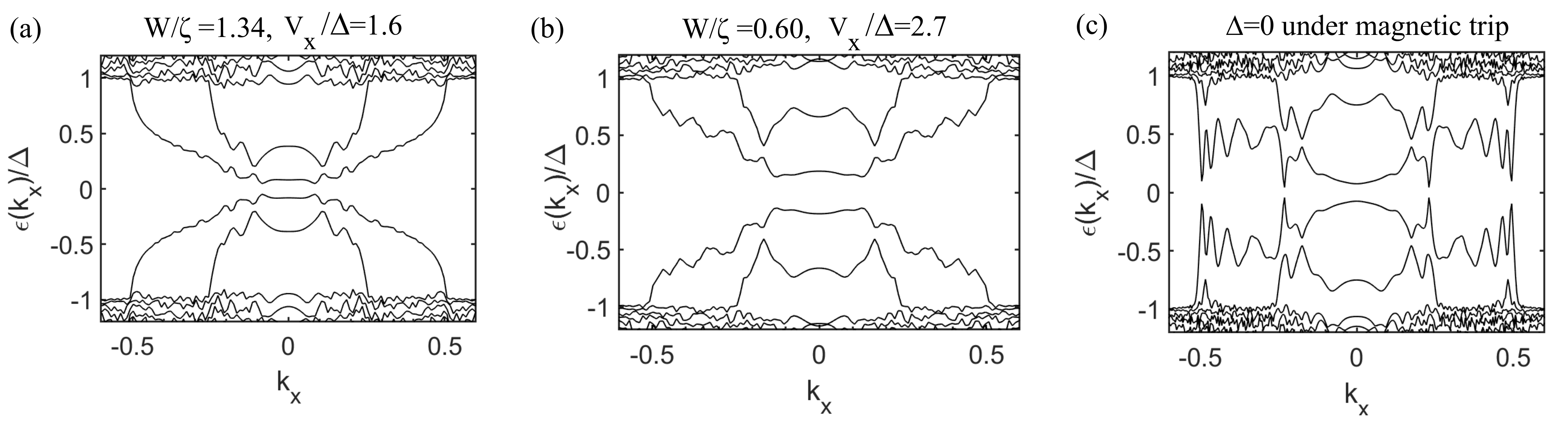}
		\caption{ (a) and (b) show the excitation spectrum $\epsilon(k_x)$ vs $k_x$ for ($W/\xi=1.34, V_x/\Delta=1.6$) and ($W/\xi=0.60, V_x/\Delta=2.7$), respectively, where there is a chemical potential step $\mu_1=100, \mu_2=20$ meV. (c) shows a typical excitation spectrum when $\Delta=0$ under the magnetic strip, i.e., being normal states. Here, $W/\xi=0.6, V_x/\Delta=1.6$.   }
		\label{fig:fig6}
	\end{figure*}
	The gap closing lines change from some isolated lines in Fig.~\ref{fig:fig5}(a) to pairs of intertwined lines with a periodic oscillation in Fig.~\ref{fig:fig5}(d) induced by the chemical potential step.  We found that the oscillation period as a function $W$ is given by $\cos(2\sqrt{m^2\alpha_R^2+2m\mu_2}W)=-1$, which gives the quantization condition
	\begin{equation}
		W=\frac{(n+\frac{1}{2})\pi}{\sqrt{m^2\alpha_R^2+2m\mu_2}}.\label{eq_node}
	\end{equation}
	In this case, we can see that the $r^2$ terms in Eq.~(\ref{gapclose}) vanish and results in Eq.~(\ref{Eq:ana2}). To show this clearly, we enlarge part of Fig.~\ref{fig:fig5}(d) as Fig.~\ref{fig:fig5}(g) and show only the analytic result as red lines.  The position of these width where $\cos(2\sqrt{m^2\alpha_R^2+2m\mu_2}W)=-1$ are highlighted as black dashed lines in Fig.~\ref{fig:fig5}(g). We note that $\sqrt{m^2\alpha_R^2+2m\mu_2} = k_{av}$ where $k_{av}=\frac{1}{2}(k_{F,\lambda}^{+}-k_{F,\lambda}^{-})$ is the average Fermi momentum over the same spin $\lambda$ orientation in the middle region according to Eq.~(\ref{momentum_middle}). Eq.~(\ref{eq_node}) can be regarded as a Bohr Sommerfeld quantization  condition where the average Fermi momentum times the width is quantized as $(n+\frac{1}{2})\pi$.
	
	
	In agreement with the result in Sec.~\ref{sec_D}, we also found the topological regime can only be accessible when the chemical potential step is present. Specifically, the regimes within a pair of  intertwined gap closing lines shown in Fig.~(\ref{gapclose}) are the topological regimes. To show this, we plot the excitation spectrum of $\epsilon(k_x=0)$ as a function of $V_x$ in Fig.~\ref{fig:fig5}(e) and Fig.~\ref{fig:fig5}(f) with a junction width $W/\xi=0.60$ and  $W/\xi=1.34$ respectively. It can be seen that the regimes within  a pair of   intertwined  lines shown in Fig.~\ref{fig:fig5}(d) are the regimes appearing after an odd number of gap closings, which manifest as topological regimes. On the contrary,  when the chemical potential step is removed, the topological regimes shrink into points, i.e., topological regimes vanish, as shown in Fig.~\ref{fig:fig5}(b) and Fig.~\ref{fig:fig5}(c). 
	
	From our derivation, the key difference between with and without chemical potential steps  is the strength of normal reflection (see Eq.~(\ref{normal_reflection})). A chemical potential step enables the normal reflection to be finite such that the gap closing lines from  Eq.~\ref{gapclose} can behave as the intertwined lines as shown in Fig.~\ref{fig:fig5}(d). Physically, the Andreev reflection,  which can only happen within intra-Rashba Fermi circle due to spin-momentum locking,  can trap the bound states and  the trivial excitation gap can be closed at finite $V_x$. However, without chemical potential steps, the excitation gap  of bound states arising from the inner and outer Rashba Fermi circle close at a same $V_x$ as shown in Fig.~\ref{fig:fig5}(b) and (c). The presence of finite normal reflections would mix the states of inner and outer Rashba Fermi circle as seen from the scattering matrix Eq.~(\ref{scatter_ma}) or Fig.~\ref{fig:fig4}(c). As a result, the gap closing lines are shifted into pairs of periodically oscillating   lines so that the topological regimes become accessible. 
	
	Moreover, as we pointed out the topological regimes exhibit periodic  oscillations as a function of $W$ setting by $\cos(2\sqrt{m^2\alpha_R^2+2m\mu_2}W)=-1$. The coefficient of $W$  is  $2k_{av}$ where $k_{av}$ is the average Fermi wavelength  of the middle region as defined above.   Hence, the chemical potential $\mu_2$ in the middle region should be comparable to the Rashba energy scale (several $m\alpha_R^2$, see Fig.~\ref{fig:fig4}(a)) and cannot be too large, otherwise the topological regime will oscillate rapidly with $W$. For example. using the parameter of gold surface states, the oscillation period as a function of $W$, i.e.,  $\pi/\sqrt{m^2\alpha_R^2+2m\mu_2}$ is about $8.4$ nm for $\mu_2=20$ meV and is reduced to  $1.8$ nm for $\mu_2=500$ meV. On the other hand, $\mu_2$ cannot be too small, because the Fermi velocity becomes small, leading to a small coherence length $\xi$. By the $W/\xi$ scaling this may require a  width $W$ which is too small to be fabricated. 
	
	Finally, we show  the features of chemical potential dependence of topological regimes from  Eq.~(\ref{gapclose}) by plotting the gap closing lines as a function  $\mu_2$ and $V_x$. These results are shown in Fig.~\ref{fig:fig5}(h),  and Fig.~\ref{fig:fig5}(i) is a zoomed-in of Fig.~\ref{fig:fig5}(h)  near $\mu_2=20$ to $\mu_2=40$ meV.  Notably, the features of  chemical potential dependence of topological regimes in  Fig.~\ref{fig:fig5}(h),  and Fig.~\ref{fig:fig5}(i) are consistent with   Fig.~\ref{fig:FIG3}(b) and Fig.~\ref{fig:FIG2}(a), respectively. Importantly, in Fig.~\ref{fig:fig5}(i), we highlighted that the topological regimes versus $\mu$ follows  the $2k_{av}$ oscillations given by  $\cos(2\sqrt{m^2\alpha_R^2+2m\mu_2}W)=-1$ (see black dashed lines).

	\subsection{Energy gap at finite $k_x$}
	Beyond the energy gap at $k_x=0$, we next look at the gap at finite $k_x$, which is also crucial for protecting the topological superconductivity. In Fig.~\ref{fig:fig6}(a) and Fig.~\ref{fig:fig6}(b), we plot the energy $E$ as a function of $k_x$ within the topological regime for ($W/\xi=1.34$, $V_x/\Delta=1.6$) and ($W/\xi=0.60, V_x/\Delta=2.7$), respectively. As expected, the increasing of junction width, i.e., the width of magnetic strip, would  decrease the excitation gap of Andreev bound states.  In practice, the width $W$ should be comparable or less than the coherence $\xi$ to obtain a sizable gap. According to our calculation,   the topological gap can be sizable  $0.1\sim 0.2\Delta$ when $W$ is reduced to be around $\xi$, which is about $60$ nm with $\mu_2=20$ meV. This estimation is consistent with the size of the EuS island used in the experiment \cite{Manna2019}. 
	
	Note that in both Fig.~\ref{fig:fig6}(a) and Fig.~\ref{fig:fig6}(b), the smallest gap at  finite $k_x$ is comparable to the one at $k_x=0$.
	This  is sharp contrast from the excitation spectrum given in  the previously studied topological superconductivity of planar Josephson junctions (see Fig.~7 of \cite{Pientka2017}), where the energy gap at some finite $k_x$ would typically be much smaller than the one at $k_x=0$. A crucial difference of the junction considered in ref.~\cite{Pientka2017} from Fig.~\ref{fig:fig4}(d) is that   there is no pairing potential in the middle part. As pointed out in ref.~\cite{Papaj2021}, the pairing potential in the middle region under the magnetic strip actually can help to form a sizable gap at finite $k_x$. To show this, we artificially turn off the pairing potential under the magnetic trip and the excitation spectrum typically behaves as Fig.~\ref{fig:fig6}(c), where the gap suddenly drops to close to zero near $k_x=\pm 2.5$ and $k_x= \pm5$, in a way that is very similar to  Fig.~7 of \cite{Pientka2017}. Therefore, the  pairing potential under the magnetic strip in our setup enables the system to avoid the problem of small energy gap at finite $k_x$. However, we note that the gap at finite $k_x$ would eventually be suppressed  by a large $V_x$,  such  as $V_x/\Delta\sim 3.9$ for $W/\xi\sim 0.60$   as shown in Supplementary Material \cite{Supp}.  This means that the maximum topological gap is in general not given by the gap at $k_x=0$ but can be smaller. We remark that this feature is in common with the model of Papaj and Fu \cite{Papaj2021}, provided the same step potential model is used.

	\section{ Conclusion and discussions} 
	In conclusion, in this work we have provided a clear understanding of the topological regimes of the EuS/Au/superconductor heterostructure. In order to put our setup in context of other setups and implementations, we next present a classification of various proposals of quasi-one-dimensional systems that are potentially scalable. The basic idea is to proximity couple a conventional SC to the surface state of a topological insulator or replace the topological insulator with a 2D semiconductor \cite{Jaysau2010,Oreg2010} or metal \cite{Patrick2010} with strong Rashba spin orbit coupling. Some form of time reversal symmetry breaking is required, which may be supplied by a magnetic field or by other means. We shall refer to these two classes as TI type or Rashba type. Next, we classify the device geometries into three types:
	
	1.	The “Nanowire geometry” consists of a narrow strip of conductor (either TI or Rashba type) sitting on top of a conventional superconductor. A lot of work has been done using semiconductor nano-crystals made with InSb or InAs \cite{Mourik2012,Albrecht2016} .  While these are not scalable, there are recent advances where the nanowire is formed lithographically in an InAs/Al heterostructure which is potentially scalable, even though significant challenges remain. In this case, the Al superconductor forms a narrow strip and the semiconductor is depleted outside of the strip to form a quasi-1D structure which can potentially support MBS \cite{nichele2017scaling}.  Another example of the nanowire geometry is the proposal of Potter and Lee, \cite{Patrick2010} who suggested the deposition of a narrow gold film on top of a conventional SC, and utilizes the surface state on the Au(111) as the active conducting channel. 
	
	2.	The “Josephson geometry”. A gap is formed between conventional SCs deposited on either TI or on Rashba semiconductors \cite{Pientka2017}. The phase of the SC on each side is separately controlled which provides the needed time reversal symmetry breaking. The advantage is that the external magnetic field can be avoided. This geometry has been realized using HgTe quantum well combined with Al superconductor \cite{Ren2019} and in InAs \cite{Fornieri2019}. However, so far the width of the junction gap is relatively wide (600 nm) \cite{Ren2019} so that a large number of conducting channels are involved and there are many in-gap states which may have obscured the possible MBS discrete level.
	
	3.	The “Ferromagnetic strip geometry”. This lies at the heart of the current paper. A narrow strip of ferromagnetic insulator such as EuS is deposited on top of a Rashba metal \cite{Patrick2010} or a topological insulator \cite{Papaj2021}, which is proximity coupled to a conventional SC. The latter is illustrated in Fig. 1(a). Time reversal symmetry breaking is provided by the exchange field of the ferromagnet, and strong external magnetic field in principle is not required.
	
	It is noteworthy that the magnetic islands have also been used to engineer two-dimensional topological superconductors \cite{Menard2017,Menard2019, Palacio2019}. However, the geometry is essentially different from ours.  The  ferromagnetic magnetic order is perpendicular to the island plane, i.e. the magnetization is out-of-plane,  and the external field that aligns the magnetization to in-plane direction is absent. This results in  chiral Majorana edge modes localizing at the boundary of magnetic islands when $V_z>\sqrt{\Delta^2+\mu^2}$, where $V_z$ is the Zeeman energy from out-of-plane magnetization.  In contrast, our setup is  a magnetic strip/Rashba superconductor heterostructure with in-plane magnetization, which is used to create MBS instead of chiral Majorana fermions. Also our work deals with  D class gapped topological superconductivity so that it is distinct from the nodal superconductivity from magnetic islands coupled Ising superconducting background considered in ref.~\cite{Godzik_2020}. 
	
	Next, we discuss the relative merits and drawbacks of the three geometries. For the nanowire, the bulk of the experimental work up to now utilizes semiconductors such as InSb or InAs. Due to the small effective mass, the Fermi momentum is small. The relevant wave-functions have long wavelengths and can be subject to manipulation by gates, and decades of experience working with gated nanostructures can be brought to bear on this system. In particular, the system can be brought to the lowest transverse sub-band created by the lateral confinement. The downside is that the slowly varying potential of the gates easily leads to possible false signatures for MBS. For example, a slowly varying tunnel potential is known to create “quasi-Majorana” which couples strongly only to one lead and looks indistinguishable from a true MBS as far as local probes such as zero bias conductance peaks are concerned \cite{kells2012near}. It is also possible to create quantum dots with trapped Andreev bound states near the junction which mimic MBS.\cite{valentini2021nontopological}  On the other hand, if the semiconductor is replaced by a metallic surface state \cite{Patrick2010} the Fermi wavelength is small and many transverse sub-bands are involved, which reduces the topological gap, as already mentioned in the introduction.
	
	For the Josephson junction geometry, the advantages are that an external magnetic field is not required. The difficulty is that the quasi-particle gap is very small for states moving along the junction, because these states are not efficiently Andreev scattered by the superconductors to receive an induced gap. We should mention that proposals have been made to alleviate it by introducing disorder scattering \cite{haim2019benefits} or kinks in the superconductor slit \cite{lesser2021phase} but these ideas remains to be tested in actual settings.
	
	The ferromagnetic  geometry proposed by us and by  Papaj and Fu  \cite{Papaj2021}  share a number of advantageous features. The optimal width of the strip is set by the superconducting coherence length under the strip, in contrast to the nanowire case, and can be relative large. The TI case has the added advantage that the topological regime is indenpent on the chemical potential, a common feature of using the surface states of topological insulators as the active conductor. In the Rashba case, the topological regime oscillates as a function of chemical potential and wire width, as shown in Fig.~\ref{fig:fig5}. On the other hand, compared with the Josephson geometry, the advantage is that the gap at large momentum along the strip does not have to be small. This is because the strip is sitting on top of a superconductor and can directly inherit pairing from it. Thus the ferromagnetic strip geometry enjoys the advantages of the other two geometries and avoid some of the key disadvantages. This is why we think this is a promising direction for future MBS research.

	\section*{Acknowledgments}
	We thank Michal Papaj annd Liang Fu for bringing their work to our attention and for discussions.  KTL acknowledges the support of the Ministry of Science and Technology of China and the HKRGC through grants MOST20SC04, RFS2021-6S03, AoE/P-701/20-2, C6025-19G, 16310219, 16309718 and 16310520.
	PAL acknowledges support by U. S. Department of Energy, Basic energy Science, under grant  DE-FG02-03ER46076, John Templeton Foundation Grants No. 39944 and 60148.
	\bibliographystyle{apsrev4-1} 
	\bibliography{Reference}

		\onecolumngrid
		\clearpage
\begin{center}
		\textbf{\large Supplementary Material for  \lq\lq{} Topological superconductivity in EuS/Au/superconductor heterostructures\rq\rq{}}\\[.2cm]
		Ying-Ming Xie$^1$, K. T. Law$^1$, Patrick A. Lee$^{2,*}$\\[.1cm]
		{\itshape ${}^1$Department of Physics, Hong Kong University of Science and Technology, Clear Water Bay, Hong Kong, China}\\
		{\itshape ${}^2$Department of Physics, Massachusetts Institute of Technology, Cambridge MA 02139, USA}\\[1cm]
\end{center}
	
	\maketitle

\setcounter{equation}{0}
\setcounter{section}{0}
\setcounter{figure}{0}
\setcounter{table}{0}
\setcounter{page}{1}
\renewcommand{\theequation}{S\arabic{equation}}
\renewcommand{\thesection}{ \Roman{section}}

\renewcommand{\thefigure}{S\arabic{figure}}
\renewcommand{\thetable}{\arabic{table}}
\renewcommand{\tablename}{Supplementary Table}

\renewcommand{\bibnumfmt}[1]{[S#1]}
\renewcommand{\citenumfont}[1]{#1}
\makeatletter

\maketitle
%
%

	\section{Details for determining topological regimes from the lattice Green's function method}
\subsection{Proximity gap}

\begin{figure}[h]
	\centering
	\includegraphics[width=0.5\linewidth]{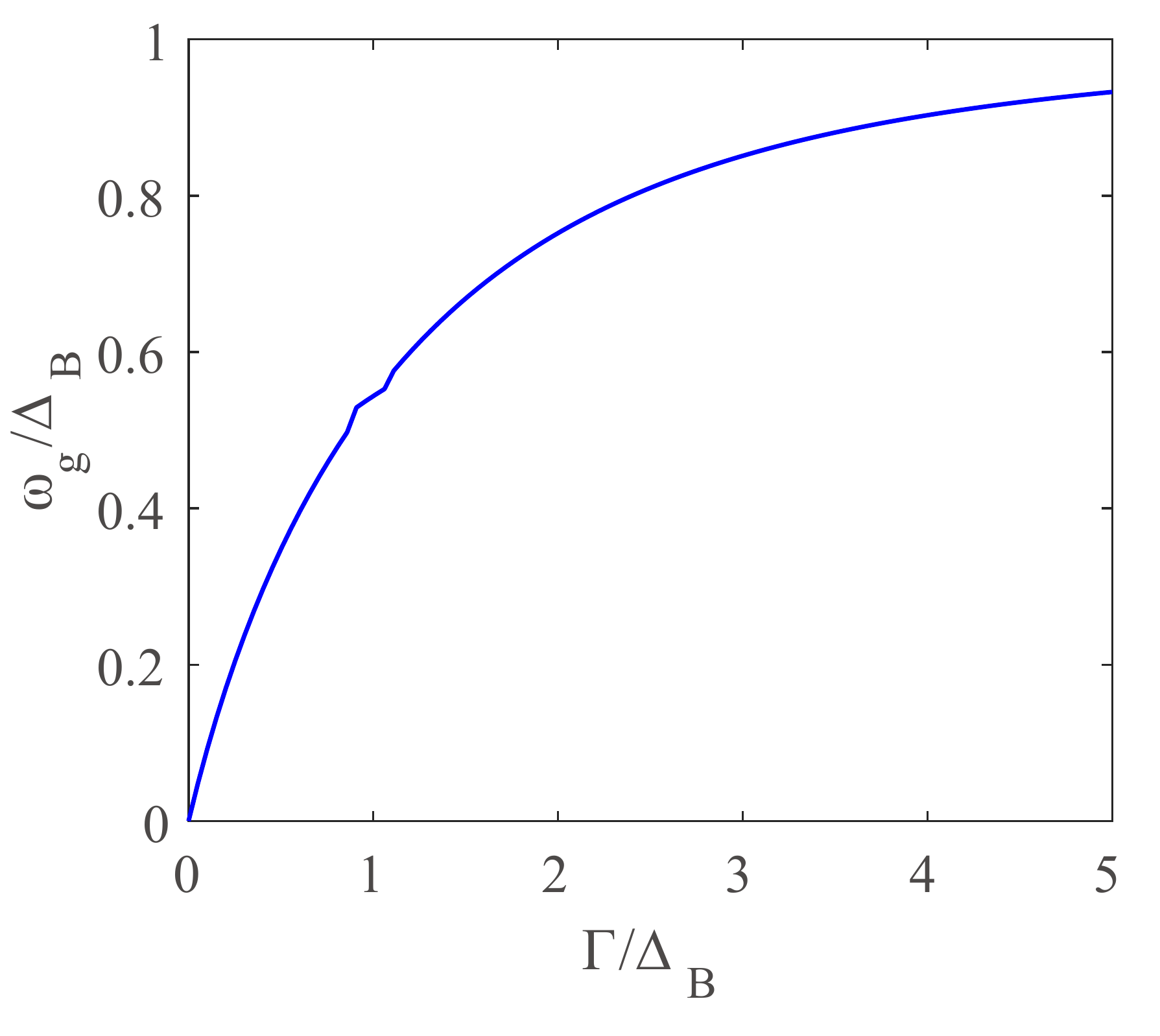}
	\caption{ The proximity gap $\omega_g$ as a function of the coupling strength $\Gamma$ (in units of $\Delta_B$), which is obtained from numerically solving Eq.~(\ref{eq: proximity_gap}) (the trivial solution $\omega_g=\Delta_B$ is dropped).}
	\label{fig:figs1}
\end{figure}

The proximity gap is given by the smallest poles of Green's function $G_0(\omega, \bm{k})$ (see the main text Eq.~(5)):
\begin{equation}
	\text{Det}(\omega_g-(1-Z(\omega_g))\Delta_B\tau_x)=0,
\end{equation}
where $\omega_g$ denotes the size of the proximity gap, $\Delta_B$ denotes the superconducting gap of the background superconductor. We obtain $\omega_g/\Delta_B=1-Z(\omega_g)$. Further substitute $Z(\omega)=\frac{1}{1+\Gamma/\sqrt{\Delta_B^2-\omega^{2}}}$,  it becomes
\begin{equation}
	\frac{\omega_g}{\Delta_B}=\frac{\Gamma}{\sqrt{\Delta_B^2-\omega^2}+\Gamma}.\label{eq: proximity_gap}
\end{equation}
In the weak and strong coupling limit, approximately, it can be found the proximity gap
\begin{equation}
	\omega_g\approx\begin{cases}
		\Gamma& \text{when } \Gamma/\Delta_B \ll 1,\\
		(1-\frac{2\Delta_B^2}{\Gamma^2})\Delta_B&   \text{when } \Gamma/\Delta_B \gg 1.
	\end{cases}
\end{equation}
The numerical solutions of Eq.~(\ref{eq: proximity_gap}) are plotted in Fig.~\ref{fig:figs1}. In the experiment, the proximity gap noto the gold surface states  is about $0.8\sim0.9\Delta_B$. Hence, from  Fig.~\ref{fig:figs1},  $\Gamma\approx 3\Delta_B$ is a good estimation of the coupling strength.  
\subsection{The effective Hamiltonian and self-energy terms  from recursive Green's function method
}

There are three segments: the upper bare gold surface  region $y\in  (W/2,+\infty)$, the middle EuS covered region $y\in(-W/2, W/2)$, the lower bare gold surface region $y\in (  -W/2,-\infty)$. The effective Hamiltonian is obtained by integrating out the bare gold surface region as a self-energy term. With Dyson equations,
\begin{equation}
	G(\omega,k_x)=(G^{-1}_0(\omega,k_x)-\Sigma_U(\omega,k_x)-\Sigma_L(\omega,k_x))^{-1}, \label{Eq:B1}
\end{equation}
where $G_0(\omega,k_x)$ is the Green's function for the EuS covered gold surface:
\begin{equation}
	G_0=\frac{Z}{(\omega^+-V_{EuS}\sigma^{x})\tau_0-Zh_{k_x}(y\in M)\tau_z-(1-Z)\Delta_B\tau_x}. \label{Eq:B2}
\end{equation}
Here $y\in M$ denotes the middle region $y\in(-W/2, W/2)$ and 
\begin{equation}
	h_{k_x}(y\in M)= 1_{N\times N}\otimes h(\omega,k_x)+\text{diag}(1_{N-1\times 1 },1)\otimes \hat{V}_c+h.c.
\end{equation}
with $N$ as the number of sites characterizing the width of EuS covered region and
\begin{eqnarray}
	&h(\omega,k_x)=(4t-\mu_2-2t\cos k_x)+\alpha_R\sin k_x\sigma^y\nonumber\\&+(1+\frac{\Gamma}{\sqrt{\Delta_B^2-\omega^2}})V_{EuS}\tau_0+\frac{\Gamma}{\sqrt{\Delta_B^2-\omega^2}}\Delta_B\tau_x,
\end{eqnarray}
and the nearest neighbor hopping matrix
\begin{equation}
	\hat{V}_c=\tau_z\otimes (-t+\frac{i}{2}\alpha_R\sigma^x).
\end{equation}

Substitute Eq.~(\ref{Eq:B2}) into Eq.~(\ref{Eq:B1}),
\begin{eqnarray}
	&G^{-1}(\omega, k_x)=Z^{-1}(\omega^+-V_{EuS}\sigma^{x})\tau_0-h_{k_x}(y\in M)\tau_z\nonumber\\&-(Z^{-1}-1)\Delta_B\tau_x-\Sigma_U(\omega,k_x)-\Sigma_L(\omega,k_x).
\end{eqnarray}
The topological regime can be solely determined by the zero frequency Hamiltonian $h_t(k_x)=-G^{-1}(\omega=0,k_x)$, which is given by 
\begin{align}
	h_t(k_x)=&h_{k_x}(y\in M)\tau_z+Z(0)^{-1}V_{EuS}\sigma_x+\nonumber\\
	&(Z(0)^{-1}-1)\Delta_B\tau_x+\Sigma_U(0,k_x)+\Sigma_L(0,k_x). \label{eff_Ha}
\end{align} 

Next, we sketch the process of evaluating the self-energy terms $\Sigma_U(0,k_x)$ and $\Sigma_L(0,k_x)$  using the recursive Green's function method.  By introducing the boundary Green's function $g_U(\omega,k_x)$ for the upper bare gold region and $g_L(\omega,k_x)$ for the lower bare gold region, the self-energy terms are written as
\begin{eqnarray}
	\Sigma_U(\omega,k_x)=\hat{V}_U^{\dagger}g_U(\omega,k_x)\hat{V}_U,\label{Eq_self1}\\
	\Sigma_L(\omega,k_x)=\hat{V}_L^{\dagger}g_L(\omega,k_x)\hat{V}_L, \label{Eq_self2}
\end{eqnarray}
where the coupling matrix $\hat{V}_U=[1,0_{1,N-1}]\otimes \hat{V}_c$, $\hat{V}_L=[0_{1,N-1},1]\otimes \hat{V}_c$ with $0_{1,N-1}$ as $1\times (N-1)$ zero matrix.
The boundary Green's function can be evaluated iteratively with
\begin{equation}
	g_{n+1,n+1}(\omega, k_x)=(\omega^+-h_{0}(\omega,k_x)-\hat{V}_c^{\dagger}g_{nn}(\omega ,k_x)\hat{V}_c),
\end{equation}
where $n$ is the column index, the intra-column Hamiltonian $h_{0}(\omega,k_x)$ can be found from the main text Eq.~(5),
\begin{eqnarray}
	&h_{0}(\omega,k_x)=(4t-\mu_1-2t\cos k_x)+\alpha_R\sin k_x\sigma^y\nonumber\\&+(1+\frac{\Gamma}{\sqrt{\Delta_B^2-\omega^2}})V_1\tau_0+\frac{\Gamma}{\sqrt{\Delta_B^2-\omega^2}}\Delta_B\tau_x.
\end{eqnarray} 
The boundary green's function $g_U(\omega,k_x)$ or $g_L(\omega,k_x)$ is given by the saturated $g_{nn}(\omega ,k_x)$ after multiple iterations. Then the self-energy terms $\Sigma_U(\omega,k_x)$ and $\Sigma_U(\omega,k_x)$ can be obtained from Eq.~(\ref{Eq_self1}) and Eq.~(\ref{Eq_self2}). Note in the numerical calculation, due to the introducing an infinitesimal imaginary part $\eta$, i.e. $\omega^{+}\equiv \omega+i\eta$, the self-energy terms $\Sigma_U(0,k_x)$ and $\Sigma_L(0,k_x)$ always contain infinitesimal imaginary parts such that $B(k_x)$ calculated later is not skew-symmetric. To fix this, these infinitesimal imaginary parts from $\eta$ needs to be removed after obtaining the self-energy terms.
Notably,  it was found the zero-frequency self-energy terms can be expanded as
\begin{equation}
	\Sigma_{U(L)} (\omega=0)\sim \tilde{\mu} \tau_z+\tilde{V}\sigma^x+\tilde{\Delta}\tau_x.
\end{equation}
These three terms physically, respectively, originate from the inhomogeneity of chemical potential,  Zeeman energy, and effective pairing potential  between the EuS covered gold surface region and bare gold surface region. It can also be seen the self-energy terms $\Sigma_{U(L)} (\omega=0)$ are real and hermitian. This is because the low energy ($\omega=0$) particles under EuS covered gold surface region only virtually enter bare gold surface region due to the presence of sizable superconducting gap.

By substituting the zero frequency self-energy terms into Eq.~(\ref{eff_Ha}), the effective Hamiltonian $h_t(k_x)$ is thus obtained. The topological invariant is calculated as
\begin{equation}
	\mathcal{M}=\text{sgn}[\text{Pf} B(k_x=0)]\times \text{sgn}[\text{Pf} B(k_x=\pi/a)] 
\end{equation}
with 
\begin{equation}
	B(k_x)=h_t(k_x)\tau_y\sigma_y.
\end{equation}

\subsection{Real-space  tight-binding Hamiltonian for Majorana wavefunction}
Here, we show the tight-binding Hamiltonian that is used to calculate the Majorana wavefunction Fig. 2(e). Since the Majorana states is closed to zero energy, i.e. $\omega\sim 0$,  we can replace $Z(\omega)$ as $Z_0=Z(\omega=0=(1+\Gamma/\Delta_B)^{-1}$ in the gold surface's Green's function $G_0(\omega, k_x)$ (Eq.~5). Comparing with the conventional form of Green's function $G=Z/(\omega^+-H)$, $Z$ the is spectral factor, the Hamiltonian that captures the Majorana states is given by
\begin{equation}
	H=Z_0h_{k_x}\tau_z+V_x\sigma^x+(1-Z_0)\Delta_B.
\end{equation}
Note here we did not further divide $H$ by a $Z_0$ factor as did in Eq.~(B7). The reason is that here  the effective Hamiltonian is not defined to characterize the topological regime, which relies on zero-frequency Green's function $-G^{-1}(\omega=0, k_x)$ only, but to study the properties of excitation states. For excitation states, it is the poles in Green's function that are essential and the poles depends on $H$ only instead of $Z^{-1}H$. We took the zero-frequency approximation, i.e. replacing $Z(\omega)$ as $Z_0=Z(\omega=0)$ and the Hamiltonian (C1) is frequency independent. The  excitation energies and wavefunctions of excitation states near zero frequency, including Majorana states, are obtained by diagonalizing Hamiltonian (C1). To obtain the real-space wavefunction for Majorana states, the Hamiltonian (C1) is written as
\begin{align}
	H&=\sum_{\bm{R}} \psi^{\dagger}(\bm{R})(Z_0(4t-\mu(\bm{R}))\tau_z+V(\bm{R})\sigma^x+\nonumber\\&+(1-Z_0)\Delta_B\tau_x) \psi(\bm{R})+\sum_{\bm{R},\bm{d}}\psi^{\dagger}(\bm{R})Z_0(-t\nonumber\\&+\frac{i}{2}\alpha_{R}(\bm{\sigma}_{\alpha\beta}\times \bm{d})\cdot\hat{\bm{z}})\tau_z\psi^{\dagger}(\bm{R}+\bm{d}).
\end{align}
Here, $\psi(\bm{R})=(c_{\uparrow}(\bm{R}),c_{\downarrow}(\bm{R}),c^{\dagger}_{\downarrow}(\bm{R}),-c^{\dagger}_{\uparrow}(\bm{R}))^{T}$ is the annihilation operator defining in Nambu basis,    $\bm{R}$ labels the positions of sites,  $V(\bm{R})=V_{EuS} (V(\bm{R})=V_1)$, $\mu(R)=\mu_2(\mu(R)=\mu_1)$ if $\bm{R}$ belongs to the EuS covered (bare gold surface) region, . $\bm{d}$ is the vector connecting the nearest neighbor sites.

We diagonalized the tight-binding Hamiltonian (C2) and plotted the wavefunction of lowest excitation energy ($\sim 4.45\times 10^{-6}$ meV) in Fig.~2(e), i.e. Majorana wavefunction, where we chose a 2000 nm$\times$ 200 nm gold surface with a 800 nm $\times$ 60 nm in the middle being covered by EuS island. Other parameters are $\mu_1=500$ meV, $V_1=0.2\Delta_B$, $\mu_2=25$ meV, $V_{EuS}=1.5\Delta_B$.

\begin{figure}
	\centering
	\includegraphics[width=0.5\linewidth]{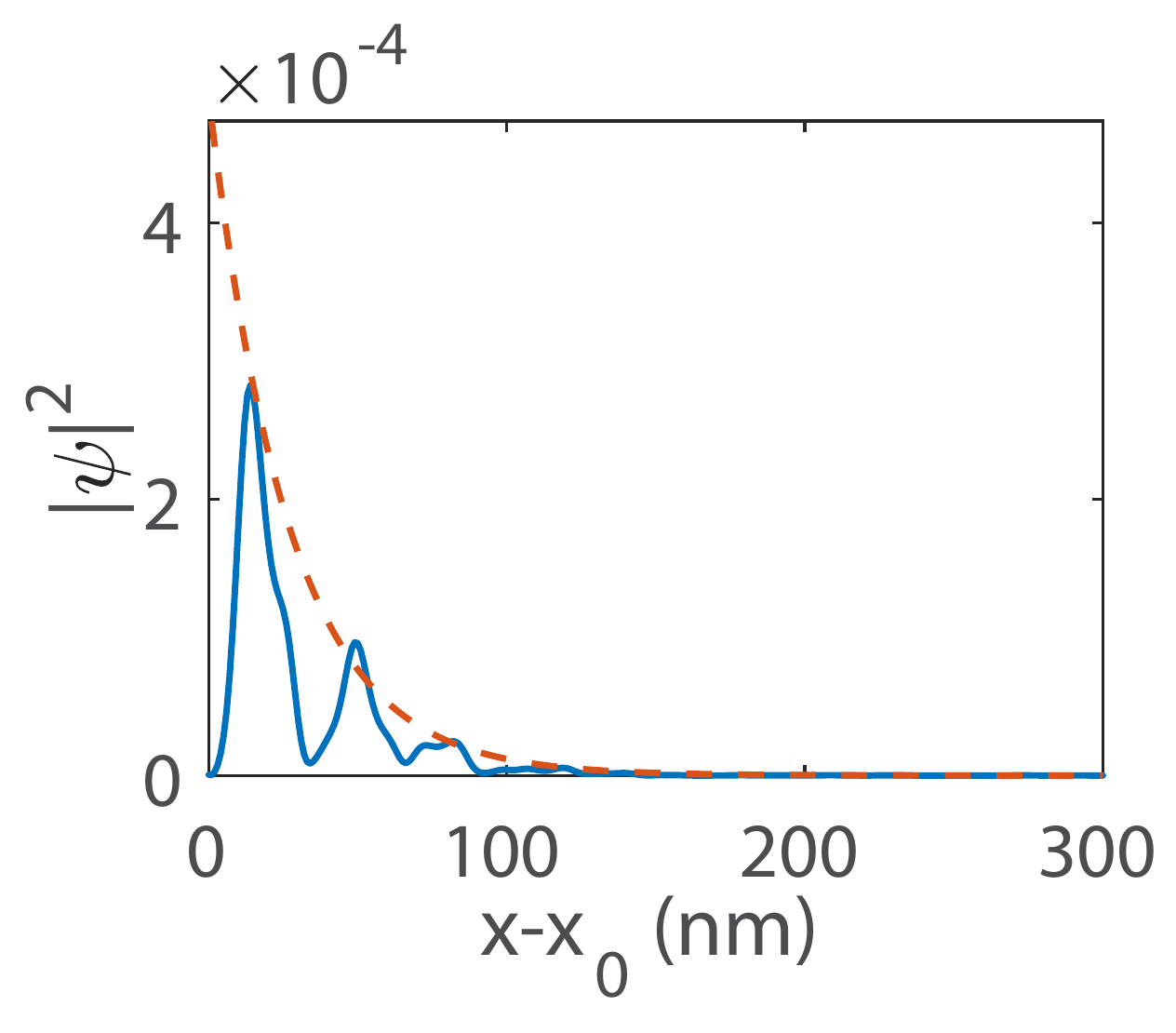}
	\caption{The exponential behaviour of Majorana wavefunction near one end of the EuS strip, where $x_0=500$ nm labels the position of the EuS strip end and only a line cut of Majorana wavefunction in Fig. 2(e)  is shown, i.e., $y$ is fixed at the middle  of the strip.}
	\label{fig:figs2}
\end{figure}
The exponential behaviour of Majorana wavefunction near one end of the EuS strip is shown in Fig. R4. The Majorana wavefunction are expected to show a exponential decay behavior, namely $\psi e^{-x/\xi}$ or $|\psi|^2 e^{-2x/\xi}$. By fitting the exponential behaviour of Majorana wavefunction in Fig. R4  with the dashed line, it can be found $\xi\approx54$ nm. On the other hand, the estimated coherence of the gold surface states is $\xi\approx t/\Delta_B\approx320$ nm, which is very long due to the large hopping of gold surface states. Based on the heuristic  considerations given in \cite{Pengyang}, the proximity effect form the bulk superconductor would renormalized $\xi$ as  $\xi^{'}=Z\xi\approx80$ nm, where $Z$=0.25 when the coupling strength $\Gamma=3\Delta_B$. Thus, qualitatively, the exponential behavior of Majorana wavefunction in Fig. R4 matches the estimated $\xi$ from effective model, although it is a bit shorter. The possible reasons that cause the decay length of Majorana wavefunction to be shorter than the estimated one may be from the partial covered geometry, the inhomogeneity and so on.

Therefore, these self-energy terms  reduce  the localization length of the observed Majorana modes to be  smaller than the coherence length of the bulk superconductor. The self-energy renormalization effect should be common in island partially covered geometries and can affect the localization of topological boundary states significantly.


\subsection{Understanding the diamond shaped nontrivial topological regime from a potential well}

\begin{figure}
	\centering
	\includegraphics[width=0.7\linewidth]{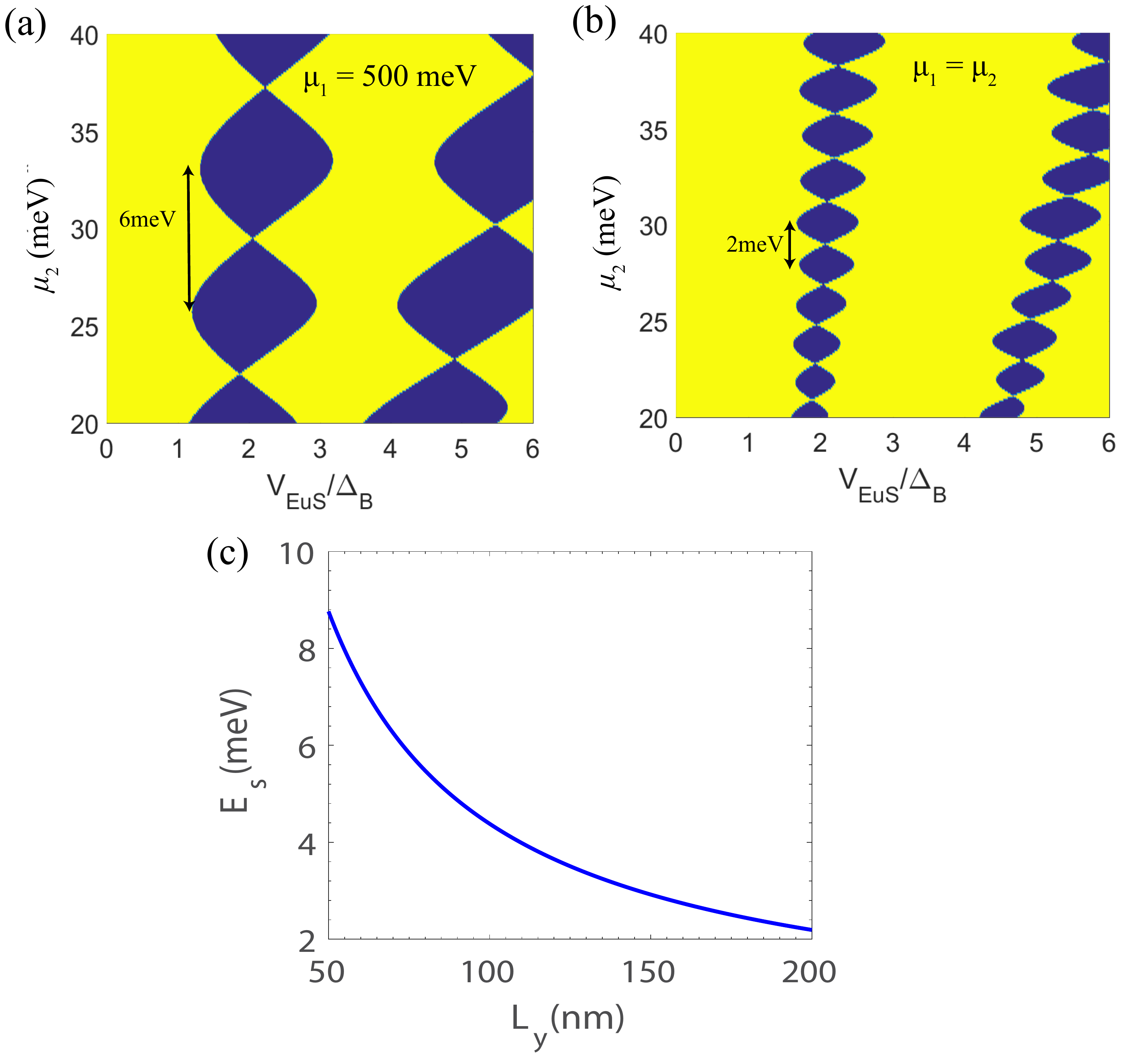}
	\caption{ (a) and (b) show the phase diagram of the heterostructure forming by the 60 nm wide EuS strip and 200nm wide gold surface. (a) is for the uniform chemical potential case where $\mu_1=\mu_2 $, (b) is for the chemical potential step case where $\mu_1$ is 500 meV. (c) shows the estimated subbands separation $E_s$ as a function of the width of wire $L_y$.}
	\label{fig:figs3}
\end{figure}

The diamond shape topological regimes in the main text Fig.~2(a) are similar to those showing in ref.~\cite{Patrick2012}. Intuitively,  the  strip EuS covered gold surface region behaves like a potential well that confines the quasiparticles inside of it.

In Fig.~\ref{fig:figs3} (a) and \ref{fig:figs3}(b), we plotted the phase diagram with 60 nm wide EuS strip and 200 nm wide gold surface with and without chemical potential step, respectively. It can be seen that the separation of  diamond shape topological regimes  is estimated as 6 meV in the presence of chemical potential step, and is reduced to about 2 meV when the chemical potential step is removed. 

Next, we understand the separation of these diamond shape topological regimes from the point view of a potential well. In a potential well with a width of $L_y$, a simple estimation of the subbands separation $E_s$ is given by 
\begin{equation}
	E_s\sim E_{N+1}-E_N=\frac{(N+1/2)\pi^2}{mL_y^2}
\end{equation}
where the band bottom energy of subbands $E_N=\pi^2N^2/2mL_y^2$, $N\sim\sqrt{2mL_y^2\mu/\pi^2}$ is  the estimated number of occupied subbands with chemical potential $\mu$.  Based on this, we plot the estimated subband separation $E_S$ as a function of $L_y$ in  Fig.~\ref{fig:figs3}(c). It can be seen from Fig.~\ref{fig:figs3}(c), a 6 meV diamond shape topological regime separation indeed corresponds to a potential well of $L_y\sim60$nm. Apparently, the scattering of the electrons by the potential step is sufficient to effectively create a potential well. On the other hand, a 2 meV diamond shape topological regime separation corresponds to a potential well of $L_y\sim200$nm, being same as the width of the whole gold surface. In other words, when the chemical potential step is removed, the separation of diamond shape topological regimes depends on the width of the whole gold surface.  This is consistent with the fact that the topological regime in the main text Fig.~3(a) almost  vanishes, because the separation of subbands is not visible for the case of a planar gold surface.

\subsection{The influence of the choice of the lattice constant $a$}

\begin{figure}
	\centering
	\includegraphics[width=0.7\linewidth]{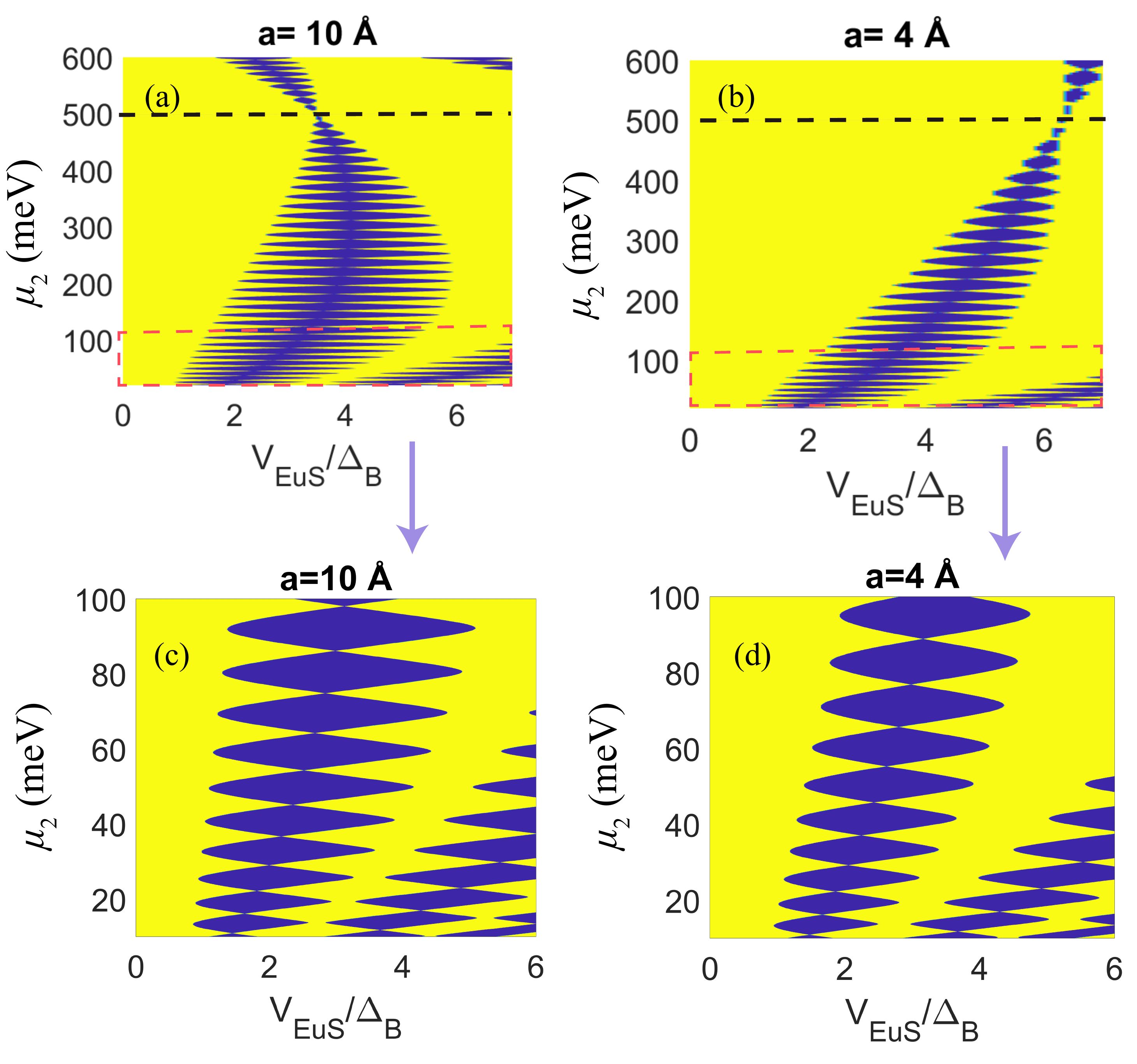}
	\caption{(a) and (b), respectively, show the topological phase diagram with a lattice constant a=10 \AA in  and a=4 \AA in  (blue color labels the topological regime with $\mathcal{M}=-1$ and yellow color labels the topological trivial region with $\mathcal{M}=1$). (c) and (d), respectively, are the zoom in topological regime of the low chemical potential region (0-100 meV) in (a) and (b). }
	\label{fig:figs5}
\end{figure}

We plotted the topological phase diagram over a wide chemical region (up to 600 meV) in shown in Fig.\ref{fig:figs5} (a) with a lattice constant of 4 \AA and Fig.\ref{fig:figs5}(b) with a lattice of 10 \AA. It can be seen that the topological regime within high chemical region is shifted by reducing the lattice constant $a$. This is because the electronic structures in the high filling region is sensitive to the lattice constant $a$. In contrast, the topological regime within the low chemical region is insensitive to the lattice constant. This is clearly seen from  Fig.\ref{fig:figs5} (c) and  Fig.\ref{fig:figs5} (d), where the zoom in topological regimes of Fig.\ref{fig:figs5} (a) and  Fig.\ref{fig:figs5} (b) at low chemical potential region are shown and they are roughly consistent. Although we used the topological phase diagram of high chemical potential region in Fig.~3(b), we want to emphasis our conclusion that the chemical potential step is essential for obtaining a sizable topological regime is not affected. In Fig.\ref{fig:figs5} (a) and  Fig.\ref{fig:figs5} (b),  the topological regime vanishes when the chemical potential step is removed, i.e. $\mu_2=\mu_1=500$ meV (see the position of black dashed line) . 

\section{Solve topological regimes using the scattering matrix method}
\subsection{Details of the derivation}
In the Nambu basis ($c_{\bm{k},\uparrow}, c_{\bm{k},\downarrow}, c^{\dagger}_{-\bm{k},\downarrow}, -c^{\dagger}_{-\bm{k},\uparrow}$),
the model Hamiltonian reads
\begin{equation}
	H(\bm{k})= [\xi_{\bm{k}}+\alpha_R (k_x\sigma_y-k_y\sigma_x)]\tau_z+V_x\sigma_x+\Delta\tau_x,
\end{equation}
where $\sigma_i$, $\tau_i$, respectively, operate on the spin and particle-hole space, the kinetic energy term $\xi_{\bm{k}}=\bm{k}^2/2m-\mu$, $V_x$ is the Zeeman energy and $\Delta$ is the pairing potential. 

The eigenenergies and eigenstates at $\Delta=0$ is written as
\begin{align}
	&E_{1,\pm}= \xi_{\bm{k}}\pm \sqrt{\alpha_Rk_x^2+(V_x-\alpha_Rk_y)^2}, \psi_{1,\pm}=\frac{1}{\sqrt{2}}(\mp ie^{i\alpha_{-}},1,0,0)^{T}\\
	& E_{2,\pm}=-\xi_{\bm{k}}\pm \sqrt{\alpha_Rk_x^2+(V_x+\alpha_Rk_y)^2}, \psi_{2,\pm}=\frac{1}{\sqrt{2}}(0,0,\pm ie^{-i\alpha_{+}},1)^{T},
\end{align}
where $\alpha_{\pm}=\text{Arg}(\alpha_R k_x+i(V_x\pm \alpha_Rk_y))$.	Being different from TI surface states \cite{Papaj2021}, the four states $(\psi_{1,\pm},\psi_{2,\pm})$ all are relevant  near Fermi energy. By projecting the model Hamiltonian in the space formed by $(\psi_{1,+},\psi_{2,-}, \psi_{1,-},\psi_{2,+})$, we obtain an effective Hamiltonian:
\begin{equation}
	H_{eff}(\bm{k})=\begin{pmatrix}
		\xi_{\bm{k}}+\alpha_Rk-V_xk_y/k&\Delta&0&0\\
		\Delta&-\xi_{\bm{k}}-\alpha_Rk-V_xk_y/k&0&0\\
		0&0&\xi_{\bm{k}}-\alpha_Rk+V_xk_y/k&\Delta\\
		0&0&\Delta&-\xi_{\bm{k}}+\alpha_Rk+V_xk_y/k
	\end{pmatrix},
\end{equation}
where the higher order terms in pairing terms are neglected as they are suppressed by $V_x/\mu$, $\Delta/\mu$. It is worthy noting $H_{eff}(\bm{k})$ is blocked diagonalized.
%

The exciting energy of this effective Hamiltonian as a function of $k_y$ at $k_x=0$ is plotted in the main text Fig.~4(b). The excitation spectrum  is fully gapped without the Zeeman energy, but becomes gapless when $V_x>\Delta$. Such gapless excitations result in some segment contours at $E=0$ as shown in the main text Fig.~4(c).

In the following, we try to solve the topological regime of a magnetic strip/ Rashba  superconductor heterostructure. The geometry of the junction we consider is displayed in the main text Fig.~4d. Here we consider the region with magnetic strip has a different chemical potential and larger Zeeman energy $V_x>\Delta$ due to the proximity effects from magnetic strip. As we showed in the main text, there are gapless excitations within the magnetic strip covered region. Those in-gap  excitations with energy $\epsilon<\Delta$ are expected to be trapped within the magnetic strip covered region as Andreev bound states, which can be labeled by a good quantum number $k_x$. The boundaries of topological phase transitions are determined by $\epsilon(k_x=0)=0$. Next, let us solve the energies of Andreev bound states at $k_x=0$.

To solve the Andreev bound states, we first need to obtain the eigen modes of different regions.
When $k_x=0$, the model Hamiltonian becomes 
\begin{equation}
	H(\bm{k})=(\xi_{\bm{k}}-\alpha_R k_y\sigma_x)\tau_z+V_x\sigma_x+\Delta\tau_x.
\end{equation} 
In this case, the model Hamiltonian $H(\bm{k})$ exhibits a chiral symmetry $[\sigma_x, H(\bm{k})]=0$. 	Thus, we can choose the spin quantization axis along $x$-direction, and block diagonalized the Hamiltonian as: 
\begin{equation}
	H(\bm{k})=\begin{pmatrix}
		H_{+}(\bm{k})&0\\0&H_{-}(\bm{k})
	\end{pmatrix}.
\end{equation}
where $H_{\lambda}=(\xi_{\bm{k}}-\lambda\alpha_R k_y)\tau_z+\lambda V_x+\Delta \tau_x$.  Because these two blocks do not mix, we can solve the bound states given by these two blocks separately. The eigen modes can be obtained from the eigen equations:
\begin{equation}
	\begin{pmatrix}
		\xi_{\bm{k}}-\lambda\alpha_Rk_y+\lambda V_x&\Delta\\
		\Delta&-\xi_{\bm{k}}+\lambda\alpha_Rk_y+\lambda V_x
	\end{pmatrix}\begin{pmatrix}
		c_{1,\lambda}\\c_{2,\lambda}
	\end{pmatrix}=\epsilon_{\lambda}\begin{pmatrix}
		c_{1,\lambda}\\c_{2,\lambda}
	\end{pmatrix},
\end{equation}
where $\epsilon_{\lambda}=\pm \sqrt{(\xi_{\bm{k}}-\lambda \alpha_Rk_y)^2+\Delta^2}+\lambda V_x$.

In the middle region where the magnetic strip covers, $V_x>\Delta$, we can rewrite $\xi_{\bm{k}}-\lambda\alpha_Rk_y=\lambda \beta^{e(h)}\sqrt{(\epsilon_{\lambda}-\lambda V_x)^2-\Delta^2}$, where $\beta^{e}=1$  for the electron-dominant  mode and $\beta^{h}=-1$ for the hole-dominant mode. The eigen wavefunctions for the zero-energy modes ($\epsilon_{\lambda}=0$) are
\begin{equation}
	\psi_{e(h),\lambda}^{\nu}(y)=\sqrt{\frac{\Delta}{2V_x}}\begin{pmatrix}
		e^{-\frac{1}{2}\beta^{e(h)}\text{acosh}\frac{V_x}{\Delta}}\\
		\lambda	e^{\frac{1}{2}\beta^{e(h)}\text{acosh}\frac{V_x}{\Delta}}
	\end{pmatrix}e^{ik_{e(h),\lambda}^{\nu}},
\end{equation}
where 
\begin{equation}
	k_{e(h),\lambda}^{\nu}=k_{F,\lambda}^{\nu}+\lambda\beta^{e(h)}\nu\frac{m\sqrt{V_x^2-\Delta^2}}{\sqrt{m^2\alpha_R^2+2m\mu_2}}, k_{F,\lambda}^{\nu}=\lambda m\alpha_R+\nu\sqrt{m^2\alpha_R^2+2m\mu_2}.
\end{equation}

In the top and bottom bare superconducting region, we set $V_x=0$. In this case,
we can rewrite $\xi_{\bm{k}}-\lambda\alpha_Rk_y= \beta^{e(h)}\sqrt{(\epsilon_{\lambda}-\lambda V_x)^2-\Delta^2}$. This gives the eigen wavefunction:
\begin{equation}
	\psi'_{e(h)}(y)=\frac{1}{\sqrt{2}}\begin{pmatrix}
		1\\
		e^{i\sigma^{e(h)\text{acos}\frac{\epsilon_\lambda}{\Delta}}}
	\end{pmatrix}e^{ik_yy}.
\end{equation}
At zero-energy modes ($\epsilon_{\lambda}=0$), the wavefunction is simplified as
\begin{equation}
	\psi'_{e(h),\nu}(y)=\frac{1}{\sqrt{2}}\begin{pmatrix}
		1\\
		i\beta^{e(h)}
	\end{pmatrix}e^{ik'^{\nu}_{e(h),\lambda}y},
\end{equation}
where
\begin{equation}
	k'^{\nu}_{e(h),\lambda}=k'^{\nu}_{F,\lambda}+\nu\beta^{e(h)}\frac{i\Delta}{\sqrt{m^2\alpha_R^2+2m\mu_1}}, k'^{\nu}_{F,\lambda}=\lambda m\alpha_R+\nu\sqrt{m^2\alpha_R^2+2m\mu_1}.
\end{equation}

Let us denote the wavefunction of the whole junction as
\begin{equation}
	\psi(y)=\begin{cases}
		c_{e}^{-}\varphi_ee^{ik'^{-}_{e,\lambda}y}+	c_{h}^{+}\varphi_he^{ik'^{+}_{h,\lambda}y}& \text{if } y\le-W/2\\
		a_e^{+}\chi_ee^{ik^{+}_{e,\lambda}y}+	a_e^{-}\chi_ee^{ik^{-}_{e,\lambda}y}+	b_h^{+}\chi_he^{ik^{+}_{h,\lambda}y}+	b_h^{-}\chi_he^{ik^{-}_{h,\lambda}y}& \text{if } -W/2\le y\le W/2\\	c_{e}^{+}\varphi_ee^{ik'^{+}_{e,\lambda}y}+	c_{h}^{-}\varphi_he^{ik'^{-}_{h,\lambda}y} & \text{if }  y\ge W/2
	\end{cases}
\end{equation}
where the vectors
\begin{equation}
	\varphi_e=\begin{pmatrix}
		1\\i
	\end{pmatrix}, 	\varphi_h=\begin{pmatrix}
		1\\-i
	\end{pmatrix}, \chi_e=\sqrt{\frac{\Delta}{2V_x}}\begin{pmatrix}
		e^{-\frac{\gamma}{2}}\\
		\lambda	e^{\frac{\gamma}{2}}
	\end{pmatrix}, \chi_h=\sqrt{\frac{\Delta}{2V_x}}\begin{pmatrix}
		e^{\frac{\gamma}{2}}\\
		\lambda	e^{-\frac{\gamma}{2}}
	\end{pmatrix}.
\end{equation}
Here $\gamma=\text{acosh}\frac{V_x}{\Delta}$.

Next, we match the boundary conditions and obtain the equation that gives rise to the zero-energy states $\epsilon(k_x=0)=0$, which corresponds to the topological regime. To make the physical process to be more clear, we use the scattering matrix method \cite{Beenakker1991,Pientka2017, Papaj2021}. Let us define
\begin{align}
	&c_e(L)=c_e^{-}e^{-ik'^{-}_{e,\lambda}\frac{W}{2}}, c_h(L)=c_h^+e^{-ik'^+_{h,\lambda}\frac{W}{2}}, c_e(U)=c_e^{+}e^{ik'^+_{e,\lambda}\frac{W}{2}}, c_h(U)=c_e^{+}e^{ik'^-_{h,\lambda}\frac{W}{2}};\\
	&a_{e}^{\nu}(L)=\sqrt{\frac{\Delta}{2V_x}}a_e^{\nu}e^{-ik^{\nu}_{e,\lambda}\frac{W}{2}},  b_{h}^{\nu}(L)=\sqrt{\frac{\Delta}{2V_x}}b_h^{\nu}e^{-ik^{\nu}_{e,\lambda}\frac{W}{2}}, a_{e}^{\nu}(U)=\sqrt{\frac{\Delta}{2V_x}}a_e^{\nu}e^{ik^{\nu}_{e,\lambda}\frac{W}{2}},b_{h}^{\nu}(U)=\sqrt{\frac{\Delta}{2V_x}}b_h^{\nu}e^{ik^{\nu}_{e,\lambda}\frac{W}{2}}. \label{trans}
\end{align} 

The continuity of the wavefunction and probability current (related to $\partial_y\psi(y)$) are parameterized as the following equations:
\begin{equation}
	\begin{pmatrix}
		1&1&0&0\\
		i&-i&0&0\\
		0&0&1&1\\
		0&0&i&-i
	\end{pmatrix}\begin{pmatrix}
		c_e^{-}(L)\\
		c_{h}^+(L)\\
		c_{e}^{+}(U)\\
		c_{h}^{-}(U)
	\end{pmatrix}=\begin{pmatrix}
		e^{-\frac{\gamma}{2}}&e^{\frac{\gamma}{2}}&0&0\\
		\lambda e^{\frac{\gamma}{2}}&\lambda e^{-\frac{\gamma}{2}}&0&0\\
		0&0&e^{-\frac{\gamma}{2}}&e^{\frac{\gamma}{2}}\\
		0&0&\lambda e^{\frac{\gamma}{2}}&\lambda e^{-\frac{\gamma}{2}}
	\end{pmatrix}\begin{pmatrix}
		a_{e}^{-}(L)\\b_{h}^{+}(L)\\a_{e}^+(U)\\b_{h}^{-}(U)
	\end{pmatrix}+\begin{pmatrix}
		e^{-\frac{\gamma}{2}}&e^{\frac{\gamma}{2}}&0&0\\
		\lambda e^{\frac{\gamma}{2}}&\lambda e^{-\frac{\gamma}{2}}&0&0\\
		0&0&e^{-\frac{\gamma}{2}}&e^{\frac{\gamma}{2}}\\
		0&0&\lambda e^{\frac{\gamma}{2}}&\lambda e^{-\frac{\gamma}{2}}
	\end{pmatrix}\begin{pmatrix}
		a_{e}^{+}(L)\\b_{h}^{-}(L)\\a_{e}^-(U)\\b_{h}^{+}(U)
	\end{pmatrix}\label{eq:sca1}
\end{equation}

\begin{align}
	\begin{pmatrix}
		k'^{-}_{F,\lambda}&k'^{+}_{F,\lambda}&0&0\\
		ik'^{-}_{F,\lambda}&-i	k'^{+}_{F,\lambda}&0&0\\
		0&0&k'^{+}_{F,\lambda}&k'^{-}_{F,\lambda}\\
		0&0&ik'^{+}_{F,\lambda}&-ik'^{-}_{F,\lambda}
	\end{pmatrix}\begin{pmatrix}
		c_e^{-}(L)\\
		c_{h}^+(L)\\
		c_{e}^{+}(U)\\
		c_{h}^{-}(U)
	\end{pmatrix}=\begin{pmatrix}
		k^{-}_{\lambda}	e^{-\frac{\gamma}{2}}&	k^{+}_{\lambda}e^{\frac{\gamma}{2}}&0&0\\
		\lambda 	k^{-}_{\lambda}e^{\frac{\gamma}{2}}&\lambda	k^{+}_{\lambda} e^{-\frac{\gamma}{2}}&0&0\\
		0&0&	k^{+}_{\lambda}e^{-\frac{\gamma}{2}}&	k^{-}_{\lambda}e^{\frac{\gamma}{2}}\\
		0&0&\lambda	k^{+}_{\lambda} e^{\frac{\gamma}{2}}&\lambda 	k^{-}_{\lambda} e^{-\frac{\gamma}{2}}
	\end{pmatrix}\begin{pmatrix}
		a_{e}^{-}(L)\\b_{h}^{+}(L)\\a_{e}^+(U)\\b_{h}^{-}(U)
	\end{pmatrix}\nonumber\\+\begin{pmatrix}
		k^{+}_{\lambda}e^{-\frac{\gamma}{2}}&	k^{-}_{\lambda}e^{\frac{\gamma}{2}}&0&0\\
		\lambda 	k^{+}_{\lambda} e^{\frac{\gamma}{2}}&\lambda 	k^{-}_{\lambda} e^{-\frac{\gamma}{2}}&0&0\\
		0&0&	k^{-}_{\lambda}e^{-\frac{\gamma}{2}}&	k^{+}_{\lambda}e^{\frac{\gamma}{2}}\\
		0&0&\lambda 	k^{-}_{\lambda} e^{\frac{\gamma}{2}}&\lambda 	k^{+}_{\lambda} e^{-\frac{\gamma}{2}}
	\end{pmatrix}\begin{pmatrix}
		a_{e}^{+}(L)\\b_{h}^{-}(L)\\a_{e}^-(U)\\b_{h}^{+}(U)
	\end{pmatrix}\label{eq:sca2}
\end{align}
Here we purposely decompose the right part into two parts, one is for the incoming state $\psi^{in}=(a_{e}^{-}(L),b^+_{h}(L), a^{+}_e(U), b_{h}^{-}(U))^{T}$ and the other is for the outgoing state $ \psi^{out}=(a_{e}^{+}(L),b^-_{h}(L), a^{-}_e(U), b_{h}^{+}(U))^{T}$. And we assumed $\mu\gg V_x,\Delta$ so that we only use $k'_{F,\lambda}$ and  $k_{F,\lambda}$ to characterize the momentum in Eq.~\ref{eq:sca2}.

First, according to the definition of Eq.~(\ref{trans}), we have $\psi_{in}=T\psi_{out}$,
where the transmission matrix is
\begin{equation}
	T=\begin{pmatrix}
		0&T_{LU}\\
		T_{UL}&0
	\end{pmatrix},	T_{LU}=\begin{pmatrix}
		e^{-ik^{-}_{e,\lambda}W}&0\\
		0&e^{-ik^{+}_{h,\lambda}W}
	\end{pmatrix}, 	T_{UL}=\begin{pmatrix}
		e^{ik^{+}_{e,\lambda}W}&0\\
		0&e^{ik^{-}_{h,\lambda}W}
	\end{pmatrix}.
\end{equation}

The Eq.~(\ref{eq:sca1}) and Eq.~(\ref{eq:sca2}) further requires:
\begin{eqnarray}
	M_1\psi^0=M_2\psi^{in}+M_3\psi^{out}\\
	M_4\psi^0=M_5\psi^{in}+M_6\psi^{out}.
\end{eqnarray}
The form of matrices can be obtained by matching with  Eq.~(\ref{eq:sca1}) and Eq.~(\ref{eq:sca2}).  These two equations can give a scattering matrix  $S$ with $\psi^{out}=S\psi^{in}$, where
\begin{equation}
	S=(M_4^{-1}M_6-M_1^{-1}M_3)^{-1}(M_1^{-1}M_2-M_4^{-1}M_5)=\begin{pmatrix}
		S_L&0\\
		0&S_U
	\end{pmatrix}.
\end{equation}
After some explicit calculations, we found
\begin{equation}
	S_L=S_U=\begin{pmatrix}
		r_{e}&r_{A}\\
		r_A&r_{h}	
	\end{pmatrix}=\begin{pmatrix}
		i\lambda r e^{i\phi_{\lambda}}&-\sqrt{1-r^2}e^{i\phi_{\lambda}}\\
		-\sqrt{1-r^2}e^{i\phi_{\lambda}}&	i\lambda r e^{i\phi_{\lambda}}.
	\end{pmatrix}.\label{scatter}
\end{equation}
Note $r_A$ is induced by the Andreev reflection for intra-Rashba Fermi circle, while $r_{e(h)}$ are induced by normal reflection between inter-Rashba Fermi circle.   Here $S_L$ and $S_R$ are the same due to the mirror symmetry, and we denote the normal reflection term and Andreev reflection term:
\begin{eqnarray}
	re^{i\phi_{\lambda}}=\frac{(\mu_1-\mu_2)\sinh\gamma}{-i\lambda(m\alpha_R^2+\mu_1+\mu_2)\sinh\gamma+\sqrt{(m\alpha_R^2+2\mu_1)(m\alpha_R^2+2\mu_2)}};\\
	\sqrt{1-r^2}e^{i\phi_{\lambda}}=\frac{\ \sqrt{(m\alpha_R^2+2\mu_1)(m\alpha_R^2+2\mu_2)}\text{cosh}\gamma}{-i\lambda(m\alpha_R^2+\mu_1+\mu_2)\sinh\gamma+\sqrt{(m\alpha_R^2+2\mu_1)(m\alpha_R^2+2\mu_2)}}.
\end{eqnarray}
It can be seen that $\phi_{\lambda}=-\phi_{-\lambda}$.
As $\psi^{in}=T\psi^{out}$ and $\psi^{out}=S\psi^{in}$, we have 	$\det[I-ST]=0$ with $I=\text{diag}(\mathbb{I},\mathbb{I})$ and $\mathbb{I}$ as the two by two identity matrix, which gives
\begin{equation}
	\det[\mathbb{I}-S_UT_{UL}S_LT_{LU}]=0. \label{Eq:bc}
\end{equation}
Inserting Eq.~(\ref{scatter}) into Eq.~(\ref{Eq:bc}), after some massage, it can be found 
\begin{equation}
	\det[\mathbb{I}-S_UT_{UL}S_LT_{LU}]=2e^{2i\lambda\theta+2i\phi_{\lambda}}[-1+r^2+r^2\cos(2\sqrt{m^2\alpha_R^2+2m\mu_2}W)+\cos(2\lambda\theta W-2\phi_{\lambda})].
\end{equation}
Therefore, we obtain the gap closing lines as
\begin{equation}
	\boxed{r^2\cos(2\sqrt{m^2\alpha_R^2+2m\mu_2}W)+\cos(2\lambda\theta W-2\phi_{\lambda})=1-r^2},\label{gapclose}
\end{equation}
where 
\begin{equation}
	\theta=\frac{m\sqrt{V_x^2-\Delta^2}}{\sqrt{m^2\alpha_R^2+2m\mu_2}}, r^2=\frac{(u_1-u_2)^2\sinh^2\gamma}{(m\alpha_R^2+\mu_1+\mu_2)^2\sinh^2\gamma+(m\alpha_R^2+\mu_1)(m\alpha_R^2+\mu_2)}.
\end{equation} 

\subsection{Tight-binding model for the magnetic strip/Rashba superconductor junction}

To verify the analytical derivation, as presented in the main text Fig.~5,  we numerically calculated the topological  phase transition boundaries with the following tight-binding model:
\begin{equation}
	H=\sum_{\bm{R}} c^{\dagger}_{\bm{R}}((4t-\mu(\bm{R}))\tau_z+V_x(\bm{R})\sigma_x+\Delta\tau_x)c_{\bm{R}}+c^{\dagger}_{\bm{R}}(-t\tau_z-\frac{i\alpha_R}{2}\tau_z\sigma_y)c_{\bm{R}+\hat{x}}+c^{\dagger}_{\bm{R}}(-t\tau_z+\frac{i\alpha_R}{2}\tau_z\sigma_x)c_{\bm{R}+\hat{y}}+\text{H.c.},
\end{equation}  
where the Zeeman energy $V_x(\bm{R})=V_x$ in the middle region covered by the magnetic strip and $V_x(\bm{R})=0$ in the other regions; the chemical potential $\mu(\bm{R})=\mu_2$ in the middle region, and  $\mu(\bm{R})=\mu_1$ in other regions (see the main text Fig. 4(d)).  To determine the topological phase transition boundary, we take periodic boundary condition along x-direction and evaluate the gap at $k_x=0$. The numerical results are summarized in the main text Fig.~5. The energy spectrum plots given in the main text Fig.~6 and below are also calculated with this tight-binding model.

\subsection{$V_x$ dependence of the BdG spectrum}
\begin{figure}
	\centering
	\includegraphics[width=1\linewidth]{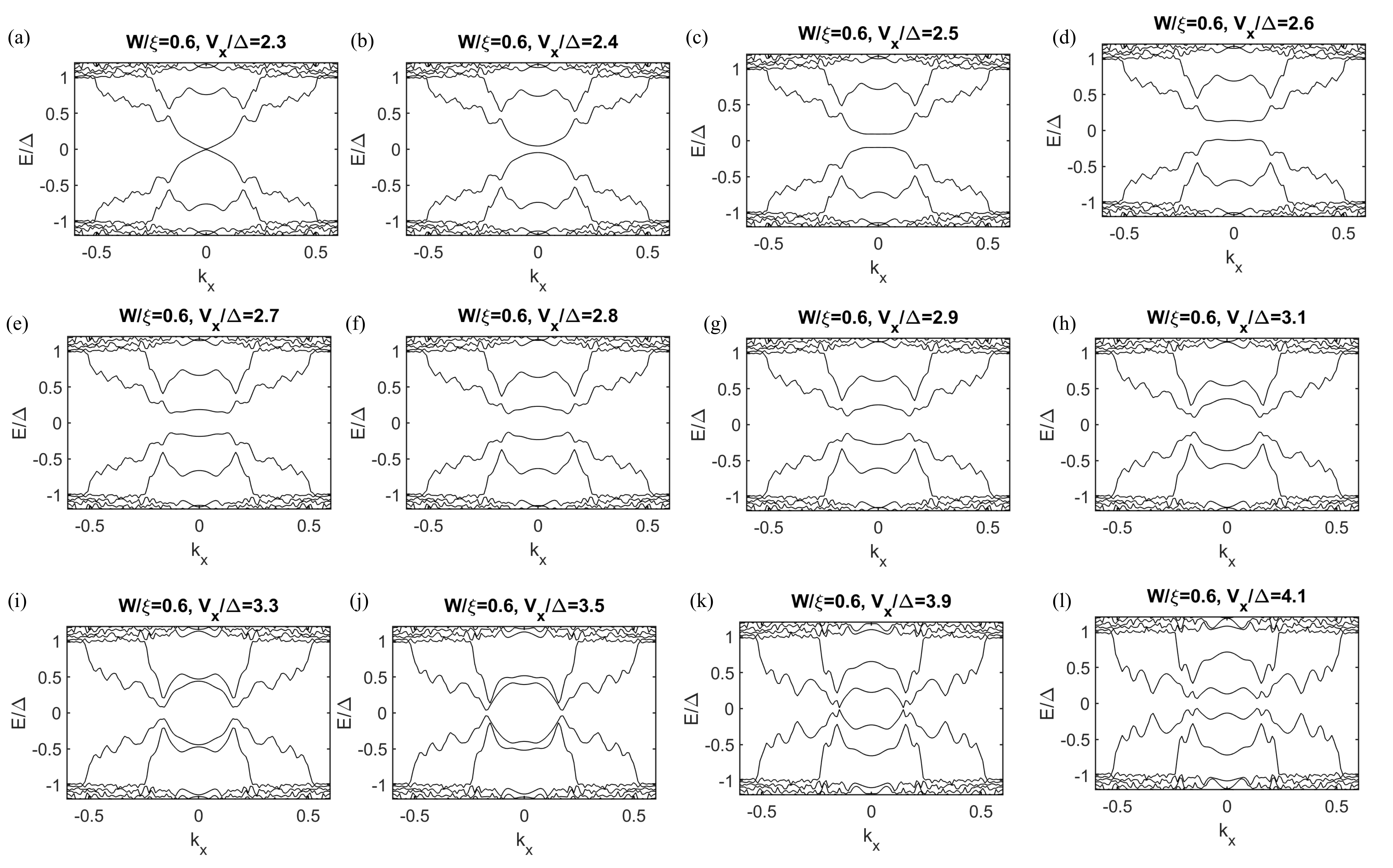}
	\caption{(a) to (l) show the BdG energy spectrum ($E$ vs $k_x$) of the magnetic strip/Rashba superconductor junction  for various Zeeman energy $V_x$ at a width $W/\xi=0.6$.  }
	\label{fig:figs6}
\end{figure}

To show the $V_x$ dependence of the BdG spectrum, we display the energy spectrum with various  $V_x$ in Fig.~\ref{fig:figs6}. Here, we fix the width to be $W/\xi=0.6$ with $\xi=v_{f2}/\Delta$. Other parameters are the same as in Fig. 5(e)  in the main text.  It can be clearly seen that  the smallest gap at finite $k_x$ is comparable to the gap at $k_x=0$ for a relatively small Zeeman energy, such as when $V_x$ is near $2.7\Delta$. This marks the maximum topological gap, which is about $0.2\Delta$  for this set of parameters. When the Zeeman energy is further increased, the gap at finite $k_x$ reduces and eventually is suppressed to be very small at about $V_x\sim3.9\Delta$, even though it does not really vanish. Note that the gap at $k_x=0$ has not yet closed and in fact the topological regime extends to $V_x=4.5\Delta$ according to Fig. 5(e). We note that the situation is very similar in the case when the Rashba metal is replaced by TI. In ref. \cite{Papaj2021} the potential under the magnetic strip was modelled by a narrow line which lives on one lattice point. If instead we employ a model where the potential is a step function which is  uniformly under the magnetic strip, similar to what is use throughout this paper, a  small gap also appears at finite $k_x$ prior to the closing of the gap at $k_x=0$.

\newpage


\end{document}